%% file: ms.tex
\documentclass[12pt,twocolumn,tighten]{aastex62}
\pdfoutput=1 %for arXiv submission
\usepackage{amsmath,amstext,amssymb}
\usepackage[T1]{fontenc}
\usepackage{apjfonts}
\usepackage[figure,figure*]{hypcap}
\usepackage{graphics,graphicx}
\usepackage{hyperref}

\received{December 22, 2018}
\revised{\today}
\accepted{---}
\submitjournal{AAS journals.}

\shortauthors{Bouma et al.}
\shorttitle{Early Arrival of WASP-4\lowercase{b}}

\listofchanges

\begin{document}

\title{WASP-4\lowercase{b} Arrived Early for the TESS Mission}

\correspondingauthor{L. G. Bouma}
\email{luke@astro.princeton.edu}

\author[0000-0002-0514-5538]{L. G. Bouma}
\affiliation{ Department of Astrophysical Sciences, Princeton
University, 4 Ivy Lane, Princeton, NJ 08540, USA}
\author[0000-0002-4265-047X]{J. N. Winn}
\affiliation{ Department of Astrophysical Sciences, Princeton
University, 4 Ivy Lane, Princeton, NJ 08540, USA}
\author[0000-0003-3438-843X]{C. Baxter}
\affiliation{ API, University of Amsterdam, P.O. Box 94249, 1090 GE
    Amsterdam, The Netherlands}
\author[0000-0002-0628-0088]{W. Bhatti}
\affiliation{ Department of Astrophysical Sciences, Princeton
University, 4 Ivy Lane, Princeton, NJ 08540, USA}
\author[0000-0002-8958-0683]{F. Dai}
\affiliation{ Department of Astrophysical Sciences, Princeton
University, 4 Ivy Lane, Princeton, NJ 08540, USA}
\author[0000-0002-6939-9211]{T. Daylan}
\altaffiliation{ Kavli Fellow }
\affiliation{ Department of Physics and Kavli Institute for Astrophysics
and Space Research, Massachusetts Institute of Technology, Cambridge, MA
02139, USA }
\author[0000-0002-0875-8401]{J.-M. D\'esert}
\affiliation{ API, University of Amsterdam, P.O. Box 94249, 1090 GE
Amsterdam, The Netherlands}
\author[0000-0002-5720-4827]{M. L. Hill}
\affiliation{Department of Earth Sciences, University of California,
Riverside, CA 92521, USA}
\author[0000-0002-7084-0529]{S. R. Kane}
\affiliation{Department of Earth Sciences, University of California,
Riverside, CA 92521, USA}
\author[0000-0002-3481-9052]{K. G. Stassun}
\affiliation{Vanderbilt University, Department of Physics \& Astronomy,
6301 Stevenson Center Lane, Nashville, TN 37235, USA}
\affiliation{Fisk University, Department of Physics, 1000 17th Avenue
N., Nashville, TN 37208, USA}
\author{J. Villasenor}
\affiliation{ Department of Physics and Kavli Institute for Astrophysics
and Space Research, Massachusetts Institute of Technology, Cambridge, MA
02139, USA }
% 
%-------------------------------------
% TESS Mission Architects:
% These authors should be listed in this order
% see https://spacebook.mit.edu/pages/viewpage.action?pageId=24543276
%-------------------------------------
%
\author{G. R. Ricker} % grr@space.mit.edu CONFIRMED
\affiliation{ Department of Physics and Kavli Institute for Astrophysics
and Space Research, Massachusetts Institute of Technology, Cambridge, MA
02139, USA }
\author[0000-0001-6763-6562]{R. Vanderspek} % roland@space.mit.edu CONFIRMED
\affiliation{ Department of Physics and Kavli Institute for Astrophysics
and Space Research, Massachusetts Institute of Technology, Cambridge, MA
02139, USA }
\author[0000-0001-9911-7388]{D. W.~Latham} % dlatham@cfa.harvard.edu CONFIRMED
\affiliation{Center for Astrophysics \textbar \ Harvard \&
Smithsonian, 60 Garden St, Cambridge, MA 02138, USA}
\author{S. Seager} % seager@mit.edu CONFIRMED
\affiliation{ Department of Earth, Atmospheric, and Planetary Sciences,
Massachusetts Institute of Technology, Cambridge, MA 02139, USA }
\author[0000-0002-4715-9460]{J. M.~Jenkins} % jon.jenkins@nasa.gov CONFIRMED
\affiliation{ NASA Ames Research Center, Moffett Field, CA 94035, USA }
%
%-------------------------------------
% 3 representatives of each of SPOC, POC, TSO, for a total of 9. 
%These 9 authors should be listed in alphabetical order
%-------------------------------------
% 3 TSO COAUTHORS
% 3 SPOC COAUTHORS
% 3 POC COAUTHORS
%
\author[0000-0002-3321-4924]{Z. Berta-Thompson} % zachory.bertathompson@colorado.EDU CONFIRMED
\affiliation{Department of Astrophysical and Planetary Sciences,
University of Colorado, Boulder, CO 80309, USA}
%
% \author[0000-0003-1963-9616]{D. A. Caldwell} % douglas.caldwell@nasa.gov % 
%%%WAITING (LIKELY YES)
% \affiliation{ NASA Ames Research Center, Moffett Field, CA 94035, USA }
% \affiliation{ SETI Institute, Mountain View, CA 94043, USA}
%
\author[0000-0001-8020-7121]{K. Col\'on} % knicole.colon@nasa.gov CONFIRMED
\affiliation{NASA Goddard Space Flight Center, Exoplanets and Stellar
Astrophysics Laboratory (Code 667), Greenbelt, MD 20771, USA}
\author{M. Fausnaugh} % faus@mit.edu CONFIRMED
\affiliation{ Department of Physics and Kavli Institute for Astrophysics
and Space Research, Massachusetts Institute of Technology, Cambridge, MA
02139, USA }
\author[0000-0002-5322-2315]{Ana Glidden} % aglidden@mit.edu CONFIRMED
\affiliation{ Department of Physics and Kavli Institute for Astrophysics
and Space Research, Massachusetts Institute of Technology, Cambridge, MA
02139, USA }
\affiliation{ Department of Earth, Atmospheric, and Planetary Sciences,
Massachusetts Institute of Technology, Cambridge, MA 02139, USA }
\author{N. Guerrero} % nmg@mit.edu CONFIRMED
\affiliation{ Department of Physics and Kavli Institute for Astrophysics
and Space Research, Massachusetts Institute of Technology, Cambridge, MA
02139, USA }
\author[0000-0001-8812-0565]{J. E. Rodriguez} % rodriguez.jr.joey@gmail.com CONFIRMED
\affiliation{Center for Astrophysics \textbar \ Harvard \&
Smithsonian, 60 Garden St, Cambridge, MA 02138, USA}
\author[0000-0002-6778-7552]{J. D. Twicken} % joseph.twicken@nasa.gov CONFIRMED
\affiliation{ NASA Ames Research Center, Moffett Field, CA 94035, USA }
\affiliation{ SETI Institute, Mountain View, CA 94043, USA}
\author{B. Wohler} % bill.wohler@nasa.gov YES
\affiliation{ NASA Ames Research Center, Moffett Field, CA 94035, USA }
\affiliation{ SETI Institute, Mountain View, CA 94043, USA}

\begin{abstract}
    The Transiting Exoplanet Survey Satellite (TESS) recently observed
    18 transits of the hot Jupiter WASP-4b.  The sequence of transits
    occurred $81.6 \pm 11.7$~seconds earlier than had been predicted,
    based on data stretching back to 2007.  This is unlikely to be the
    result of a clock error, because TESS observations of other hot
    Jupiters (WASP-6b, 18b, and 46b) are compatible with a constant
    period, ruling out an 81.6-second offset at the 6.4$\sigma$ level.
    The 1.3-day orbital period of WASP-4b appears to be decreasing at
    a rate of $\dot{P} = -12.6 \pm 1.2$ milliseconds per year.  The
    apparent period change might be caused by tidal orbital decay or
    apsidal precession, although both interpretations have
    shortcomings.  The gravitational influence of a third body is
    another possibility, though at present there is
    \replaced{no}{minimal} evidence for such a body. Further
    observations are needed to confirm and understand the timing
    variation.
\end{abstract}


%%%%%%%%%%%%%%%%%%%%%%%%%%%%%%%%%%%%%%%%%%
\section{Introduction}
\label{sec:intro}

Although the Transiting Exoplanet Survey Satellite (TESS\added{,
\citealt{ricker_transiting_2015}}) is designed to detect new planets,
it is also \replaced{performing precise photometric monitoring
of}{precisely monitoring} most of the planets that have been
discovered by ground-based transit surveys over the last two decades.
One \replaced{of the applications}{application} of the new TESS data
is to search for timing anomalies in \added{previously known }hot
Jupiter sytems.  Long-term monitoring of hot Jupiter transit and
occultation times should eventually reveal \replaced{two distinct
processes}{variations caused by three different phenomena}.

\explain{The following four paragraphs have been modified to add
emphasis to the possibility of an outer planet explaining the timing
deviations. Some background information given in the original
submission has been removed.}

First, the orbits of most hot Jupiters should shrink because of tidal
orbital decay
\citep{counselman_outcomes_1973,hut_stability_1980,levrard_falling_2009,matsumura_tidal_2010}.
Directly measuring the rate of decay might lead to an improved
understanding of how friction dissipates the energy of tidal
disturbances (a problem reviewed by \citealt{Mazeh2008} and
\citealt{ogilvie_tidal_2014}). Second, if hot Jupiter orbits are
appreciably eccentric, then long-term timing studies should reveal
rotation of the orbital ellipse within the orbital plane (``apsidal
precession'').  If this effect were observed, it could yield a measure
of the planet's Love number, which would constrain the planet's
interior structure \citep{ragozzine_probing_2009}.  The most
convincing direct evidence yet found for either orbital decay or
apsidal precession of a hot Jupiter is the case of WASP-12b, which has
a transit period that has decreased by about 30~milliseconds per year
over the last decade \citep{maciejewski_departure_2016,patra_2017}.

The final effect of interest that can produce period changes in hot
Jupiter systems is gravitational acceleration caused by massive outer
companions \citep[{\it e.g.},][Section~4]{agol_detecting_2005}.
Prototypes include WASP-53 and WASP-81, systems in which the inner hot
Jupiters are periodically perturbed by eccentric brown dwarf
companions with semimajor axes of a few astronomical units
\citep{triaud_peculiar_2017}.

Here, we present evidence for a timing anomaly in the WASP-4 system.
The hot Jupiter WASP-4b orbits a G7V star every 1.34 days,
corresponding to an orbital distance of 5.5 stellar radii
\citep{wilson_wasp-4b_2008,huitson_gemini_2017}.  It is a good target
to search for departures from a constant period, because transits have
been observed since 2007.  The orbital eccentricity is less than 0.018
(2$\sigma$), based on the work of \cite{knutson_friends_2014}, who
combined the available transit times, occultation times, and Doppler
data.  The sky projection of the stellar obliquity is also compatible
with zero, within about 10 degrees
\citep{triaud_spin-orbit_2010,sanchis-ojeda_starspots_2011}.

In what follows, \S~\ref{sec:observations} presents the new TESS
observations, and \S~\ref{sec:timing} describes our timing analysis.
We tried fitting the data with three models: a constant period; a
steadily shrinking period; and a slightly eccentric, precessing orbit.
A constant period can be ruled out\replaced{, but we}{. We} cannot
distinguish between the\replaced{ latter two possibilities}{
  possibilities of a decaying orbit, a precessing orbit, and the
  unmodeled possibility of an orbit being gravitationally perturbed by
  an outer companion}.  \replaced{Either possibility}{Any of the three
scenarios} would have interesting implications\replaced{, described in
\S~\ref{sec:implications}}{ (\S~\ref{sec:implications}), and more data
are required for a definitive ruling (\S~\ref{sec:future}).}
Appendix~\ref{sec:verify_tess} considers the possibility that the
WASP-4 timing anomaly is due to an error in timestamps in the TESS
data products.  We found this possibility to be unlikely because none
of the other hot Jupiters we examined show a timing offset with the
same amplitude as was seen for WASP-4.

%%%%%%%%%%%%%%%%%%%%%%%%%%%%%%%%%%%%%%%%%%
\section{New transits and system parameters}
\label{sec:observations}

\begin{figure*}[t]
    \begin{center}
		\includegraphics[height=0.97\textheight]{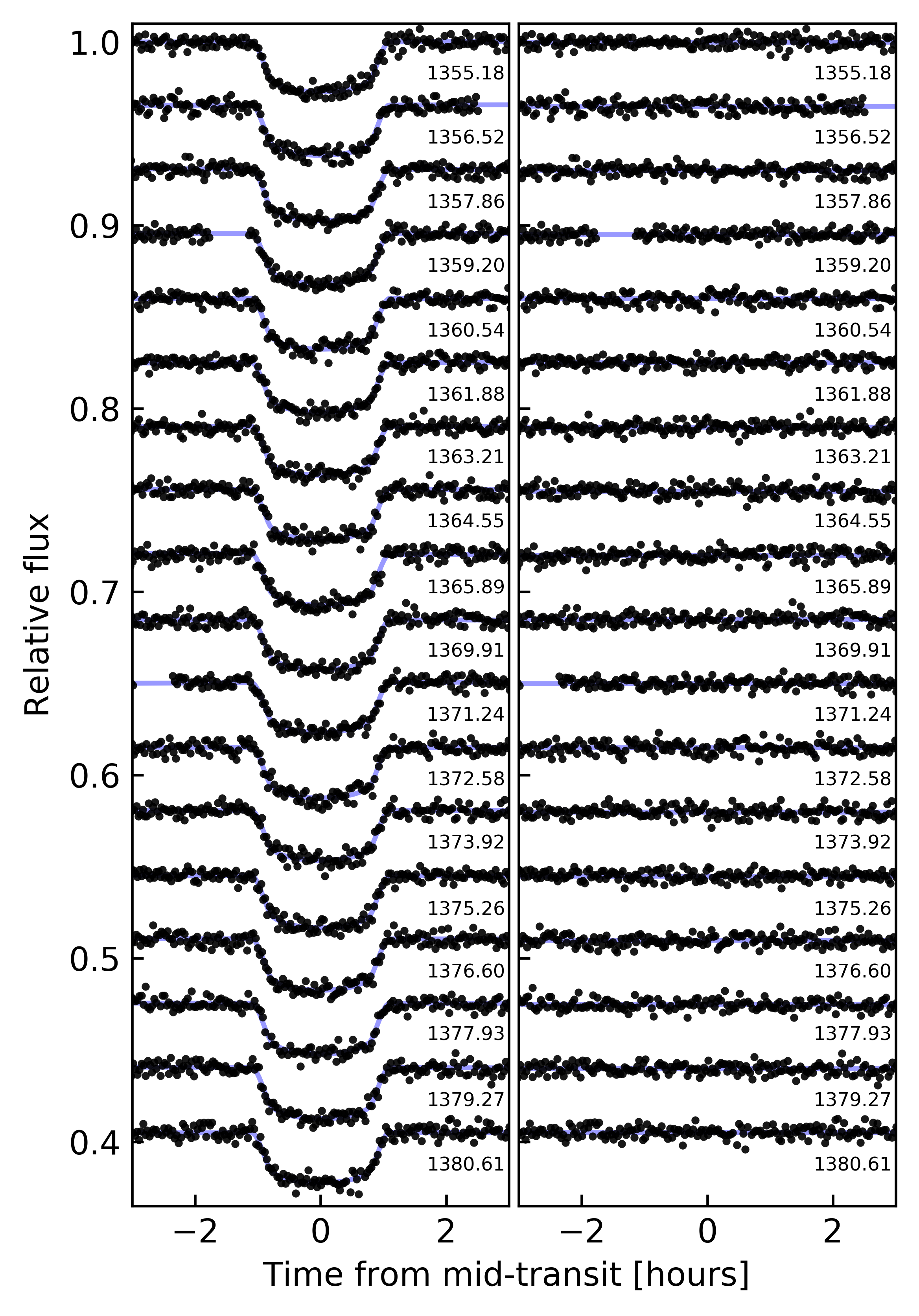}
    \end{center}
    \vspace{-0.8cm}
    \caption{
      {\bf TESS observations of WASP-4b.} On the left, black points
      are TESS flux measurements, with a vertical offset applied. Blue
      curves are best-fit models. The numbers printed next to each
      lightcurve are the approximate transit times expressed in BJD
      minus 2{,}457{,}000.  The right side shows the residuals.
       \label{fig:lightcurves}
    }
\end{figure*}

\begin{figure*}[t]
    \begin{center}
        \includegraphics[width=0.7\textwidth]{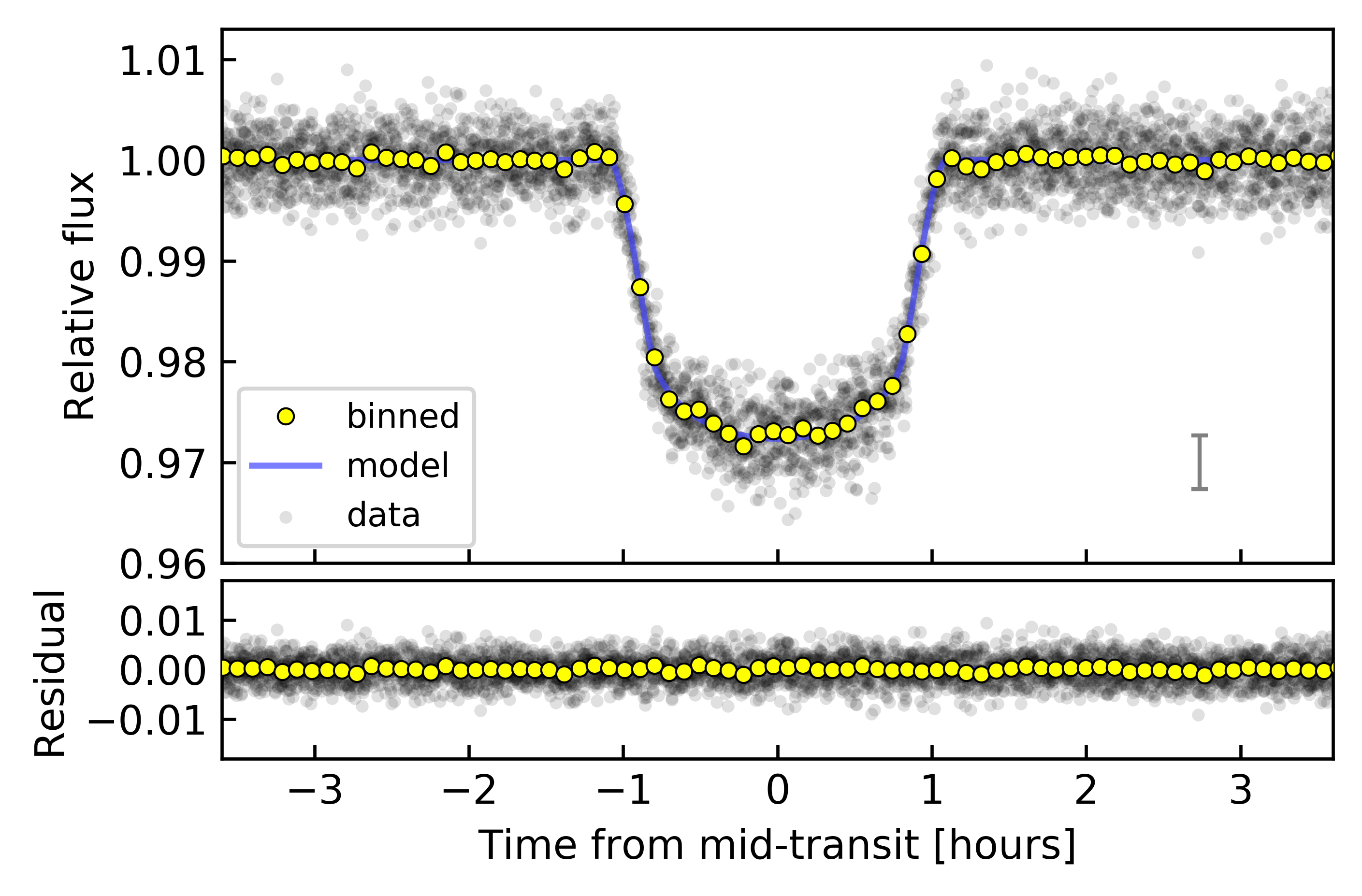}
    \end{center}
    \vspace{-0.7cm}
    \caption{
        {\bf Phase-folded lightcurve of WASP-4b.} Gray points are
        TESS \replaced{data}{flux measurements, with median 
        $1\sigma$ uncertainty shown in the lower 
        right.} 
        Yellow points are binned measurements.  The bottom
        panel shows the residuals.  The fit to the phase-folded
        transit (blue line) is used when measuring mid-transit times
        \replaced{for the}{of} individual transits (see 
        \S~\ref{sec:measurement}).
        \label{fig:phasefold}
    }
\end{figure*}

\subsection{Observations}

WASP-4 was observed by TESS with Camera 2 from August 23 to September
20, 2018, within the second ``sector'' of science operations.  The
star is designated as TIC 402026209 in the TESS Input Catalog
\citep{stassun_TIC_2018}.  The pixel data for an $11\times11$ array
surrounding WASP-4 were averaged into 2-minute stacks by the onboard
computer.  The data were downlinked via the Deep Space
Network\footnote{\url{deepspace.jpl.nasa.gov}}, and the
spacecraft timestamps were calibrated against the ground-station
clocks.  The spacecraft clock times were then transformed by the
Payload Operations Center into the {\it Temps Dynamique Barycentrique}
(TDB) reference system.  The images were then reduced to lightcurves
by the Science Processing Operations Center (SPOC) at NASA
Ames~\citep{jenkins_tess_2016}.  During this processing, the SPOC used
the known spacecraft trajectory to compute the barycentric time
corrections on a target by target basis, and expressed the timestamps
as Barycentric TESS Julian Dates (BTJD), which is simply the
\replaced{b}{B}arycentric Julian \replaced{d}{D}ate minus
2{,}457{,}000.  Lightcurves that were flagged\deleted{ as Threshold
Crossing Events} by the SPOC pipeline \added{as crossing a transit
detection threshold} were then vetted and released by the MIT TESS
Science Office to the Mikulski Archive for Space Telescopes on
November 29, 2018~\citep{ricker_tess_alerts_2018}.

We began our analysis with the Presearch Data Conditioning (PDC)
lightcurve\replaced{
  \citep{smith_kepler_apertures_2017,smith_kepler_PDC_2017}}{, which
has had non-astrophysical variability removed through the methods
discussed by \citet{smith_kepler_apertures_2017} and
\citet{smith_kepler_PDC_2017}}.  We \added{then} processed the
lightcurve as follows.  First, we removed all points with non-zero
quality flags.  This removed data that might have been adversely
affected by ``momentum dumps,'' the firing of thrusters and resetting
of reaction wheels\footnote{\added{The spacecraft pointing and
momentum dumps are described in the data release notes:
}\url{archive.stsci.edu/missions/tess/doc/tess_drn/tess_sector_02_drn02_v01.pdf}}
that took place every 2.5 days during sector 2\deleted{
  \citep{tess_data_product_description_2018}}.  The data during these
events were assigned quality flags corresponding to ``Reaction Wheel
Desaturation Event'' and ``Manual Exclude'' \added{\citep[][Table
28]{tess_data_product_description_2018}}.  For WASP-4, these flags
were simultaneously set for 54 distinct cadences, and there were 10
momentum dumps, averaging about 10 minutes of flagged data per dump.
Out of caution, we clipped out an additional 10 minutes before and
after every momentum dump.  We also removed the data within the first
and last hours of both orbits, because of \replaced{ramp-like
systematic effects that appear}{correlated red noise that appears}
during those time ranges. 

All told, we removed 8\% of the original data points, and were left
with 18{,}165 measurements of the relative flux of WASP-4.  We
normalized the data by dividing out the median flux.  We converted the
timestamps from BTJD$_{\rm TDB}$ into BJD$_{\rm TDB}$ by adding the
appropriate 2{,}457{,}000 day offset
\citep{tess_data_product_description_2018}.  Many of these and
subsequent processing steps were performed using
\texttt{astrobase}~\citep{bhatti_astrobase_2018}. We did not
``flatten'' the lightcurves, as is often done with splines,
polynomials, or Gaussian processes.  Instead, we modeled the
out-of-transit flux variations simultaneously with the transit
parameters, as described below.

\subsection{Measuring the transit times}
\label{sec:measurement}

Using the cleaned PDC lightcurve, we applied the Box Least Squares
algorithm \citep{kovacs_box-fitting_2002} to estimate the orbital
period, transit duration, and a reference epoch using the TESS data
alone.  Based on the results, we isolated the data within 4 transit
durations of each transit midpoint.  To find the transit parameters
that best fit the data, we first fitted a line to the out-of-transit
flux measurements surrounding each transit, and divided it out.  We
then created a phase-folded lightcurve \added{from all 18 transits}
and fitted a standard transit model using the analytic formulae given
by \citet{mandel_analytic_2002} and implemented by
\replaced{\citet{kreidberg_batman_2015}}{\citet[][\texttt{BATMAN}]{kreidberg_batman_2015}}
We assumed the orbit to be circular, consistent with the limits from
radial velocities and occultation timing
\citep{beerer_secondary_2011,knutson_friends_2014,bonomo_gaps_2017}.
The free parameters were the reference epoch, the planet to star
radius ratio $R_{\rm p}/R_\star$, the orbital distance to stellar
radius ratio $a/R_\star$, the inclination $i$, two quadratic
limb-darkening coefficients $(u_{\rm linear}, u_{\rm quad})$, and the
orbital period $P$.

We sampled the posterior probability distribution for all the
parameters using the algorithm proposed by
\citet{goodman_ensemble_2010} and implemented by
\replaced{\citet{foreman-mackey_emcee_2013}}{\citet[][\texttt{emcee}]{foreman-mackey_emcee_2013}}.
Table~1 gives the results, which are in reasonable agreement with the
parameters reported by {\it e.g.},
\citet{southworth_high-precision_2009} and
\citet{huitson_gemini_2017}.  Figure~\ref{fig:phasefold} shows the
phase-folded lightcurve.

To measure the transit times, we returned to the `cleaned' PDC time
series and fitted the data within four transit durations of each
transit separately. We used four free parameters: the time of
mid-transit $t_{\rm tra}$, the planet-to-star radius ratio, and the
slope and intercept of a linear trend to account for any slow
variations unrelated to the transit.  We fixed the remaining
parameters at the values that had been determined from the
phase-folded TESS lightcurve.  The uncertainty in each
\added{photometric} data point was set equal to the root-mean-square
(rms) level of the out-of-transit data.

To verify that the measured uncertainties are estimated accurately, we
computed the \deleted{reduced }$\chi^2$\added{ value} for a linear
ephemeris fit to the measured TESS mid-transit times.  We found that
$\chi^2 = 9.2$, with $n=16$ degrees of freedom.  The variance of the
$\chi^2$ distribution is $2n$, so we would expect $\chi^2 = 16 \pm
5.7$.  Visually inspecting the residuals showed that the error
variance had been overestimated, so we multiplied the measured TESS
errors by a factor $f=0.76$, forcing a reduced $\chi^2$ of unity.
This lowered the mean uncertainty of the transit midtimes from $29.8$
to $22.6$ seconds.  We verified that omitting this step did not
appreciably alter any of our conclusions.

Figure~\ref{fig:lightcurves} shows the lightcurve of each individual
transit, the best-fit models, and the residuals.  Table~2 reports the
mid-transit times and their uncertainties.  After binning the
residuals to 1-hour windows, the lightcurves have an rms scatter of
586\,{\rm ppm}.  The pre-launch TESS noise
model\footnote{\url{github.com/lgbouma/tnm}, commit \texttt{be06f09}}
(\citealt{winn_photonflux_2013}, \citealt{Sullivan_2015} Section 6.4)
would have predicted an error budget consisting of the following terms
added in quadrature: 410\,{\rm ppm} from photon-counting noise,
202\,{\rm ppm} from detector read noise, and 673\,{\rm ppm} from the
zodiacal background light.  The level of background light appears to
have been overestimated in the model.

\subsection{Star and planet parameters}
\label{sec:system_parameters}

\begin{deluxetable}{lccc}
\tabletypesize{\scriptsize}
\tablecaption{Selected system parameters of WASP-4b\label{tbl:params}}
\tablenum{1}

\tablehead{
\colhead{Parameter} & \colhead{Value} & \colhead{68\% Confidence Interval} & \colhead{Comment}
}

\startdata
{\it Transit/RV parameters:} & & & \\
  $R_{\rm p}/R_\star$                        & $0.15201$              & $+0.00040$, $-0.00033$      & A \\
  $i$~[deg]                                  & $89.06$                & $+0.65$, $-0.84$            & A \\
  $a/R_\star$                                & $5.451$                & $+0.023$, $-0.052$          & A \\
  $u_{\rm linear}$                           & $0.382$                & ---                         & A \\
  $u_{\rm quad}$                             & $0.210$                & ---                         & A \\
  $K$~[m~s$^{-1}$]                           & $241.1$                & $+2.8$, $-3.1$              & B \\
{\it Stellar parameters:} & & & \\
  $T_{\rm eff}$~[K]                          & $5400$                 & $\pm 90$                    & C \\
  $\log g_\star$~[cgs]                       & $4.47$                 & $\pm 0.11$                  & C \\
  $[{\rm Fe/H}]$                             & $-0.07$                & $\pm 0.19$                  & C \\
  $F_{\rm bol}$~[erg~cm$^{-2}$~s$^{-1}$]     & $2.802\times10^{-10}$  & $\pm 0.076\times10^{-10}$   & D \\
  $A_V$~[mag]                                & $0.03$                 & $+0.02, -0.01$              & D \\
  $\pi$~[mas]                                & $3.7145$               & $0.0517$                    & F \\
  $R_\star$~[R$_{\odot}$]                    & $0.893$                & $\pm 0.034$                 & E \\
  $\rho_\star$~[g~cm$^{-3}$]                 & $1.711$                & $+0.022$, $-0.048$          & E \\
  $M_\star$~[M$_{\odot}$]                    & $0.864$                & $+0.084$, $-0.090$          & E \\
  $T$ magnitude                              & $11.778$               & $\pm 0.018$                 & G \\
{\it Planetary parameters:} & & & \\
  $a$~[AU]                                   & $0.0226$               & $+0.0007$, $-0.0008$        & E \\
  $M_p$~[M$_{\rm Jup}$]                      & $1.186$                & $+0.090$, $-0.098$          & E \\
  $R_p$~[R$_{\rm Jup}$]                      & $1.321$                & $\pm 0.039$                 & E \\
\enddata

\tablecomments{
  (A) From phase-folded TESS lightcurve (\S~\ref{sec:measurement}).
  Orbital periods are in Table~4. The limb darkening parameters were
  allowed to float around the \citet{claret_limb_2017} prediction, but
  were unconstrained.
  (B) \citet{triaud_spin-orbit_2010}.
  (C) From HARPS spectra \citep{doyle_accurate_2013}.
  (D) \citet{stassun_accurate_2017}.
  (E) This work, see \S~\ref{sec:system_parameters}.
  (F) \citet{gaia_collaboration_gaia_2018}.
  (G) \citet{stassun_TIC_2018}.
}

\vspace{-1cm}
\end{deluxetable}

We calculated the stellar and planetary parameters in the following
way.  We computed the star's spectral energy distribution based on the
{\it Gaia} DR2 parallax (after making the small correction advocated
by \citealt{stassun_evidence_2018}) and the broadband magnitudes from
the available all-sky catalogs: $G$ from {\it Gaia\/} DR2, $B_T$ and
$V_T$ from {\it Tycho-2}, $BVgri$ from {\it APASS}, $JHK_S$ from {\it
2MASS}, and the {\it WISE}~1--4 passbands, thus spanning the
wavelength range 0.4--22~$\mu$m.
We adopted the effective temperature
from the work by \citet{doyle_accurate_2013}, who determined the
spectroscopic parameters of WASP-4 using high-signal-to-noise
observations with the High Accuracy Radial-velocity Planet Searcher
(HARPS).  Then, we determined the stellar radius through the
combination of the bolometric luminosity and the effective
temperature, using the Stefan-Boltzmann law.  To determine the stellar
mass, we first computed the mean stellar density based on the value of
$a/R_\star$ that gave the best fit to the phase-folded TESS lightcurve
(for the relevant equation, see \citealt{seager_unique_2003} or
\citealt{winn_exoplanet_2010}).  The mass was calculated from the
radius and density, and the orbital distance was also calculated from
the radius and $a/R_\star$.  The planetary radius was calculated as
the product of $R_\star$ and $R_{\rm p}/R_\star$.  Finally, the planet
mass was calculated based on the stellar mass, the radial-velocity
amplitude observed by \citet{triaud_spin-orbit_2010}, and the orbital
inclination.

Table~1 gives the resulting parameters, which we adopted for the
remaining analysis.  \added{The uncertainties in our derived stellar
and planetary parameters are propagated according to standard analytic
formulae, under the assumption that the variables are uncorrelated and
normally distributed.} Our \added{system} parameters are in agreement
with those of previous investigators, but have the benefit of
incorporating the {\it Gaia} parallax
\citep{wilson_wasp-4b_2008,gillon_discovery_2009,winn_transit_2009,southworth_homogeneous_2011,petrucci_no_2013,huitson_gemini_2017}.
By comparing the star's luminosity and spectroscopic parameters with
the outputs of the Yonsei-Yale stellar-evolutionary models, we found
that WASP-4 is a main-sequence star with an age of approximately
$7\,{\rm Gyr}$ \added{\citep{demarque_y2_2004}}.

%%%%%%%%%%%%%%%%%%%%%%%%%%%%%%%%%%%%%%%%%%
\section{Timing analysis}
\label{sec:timing}

\begin{figure}[t]
    \begin{center}
        \leavevmode
        \includegraphics[width=0.49\textwidth]{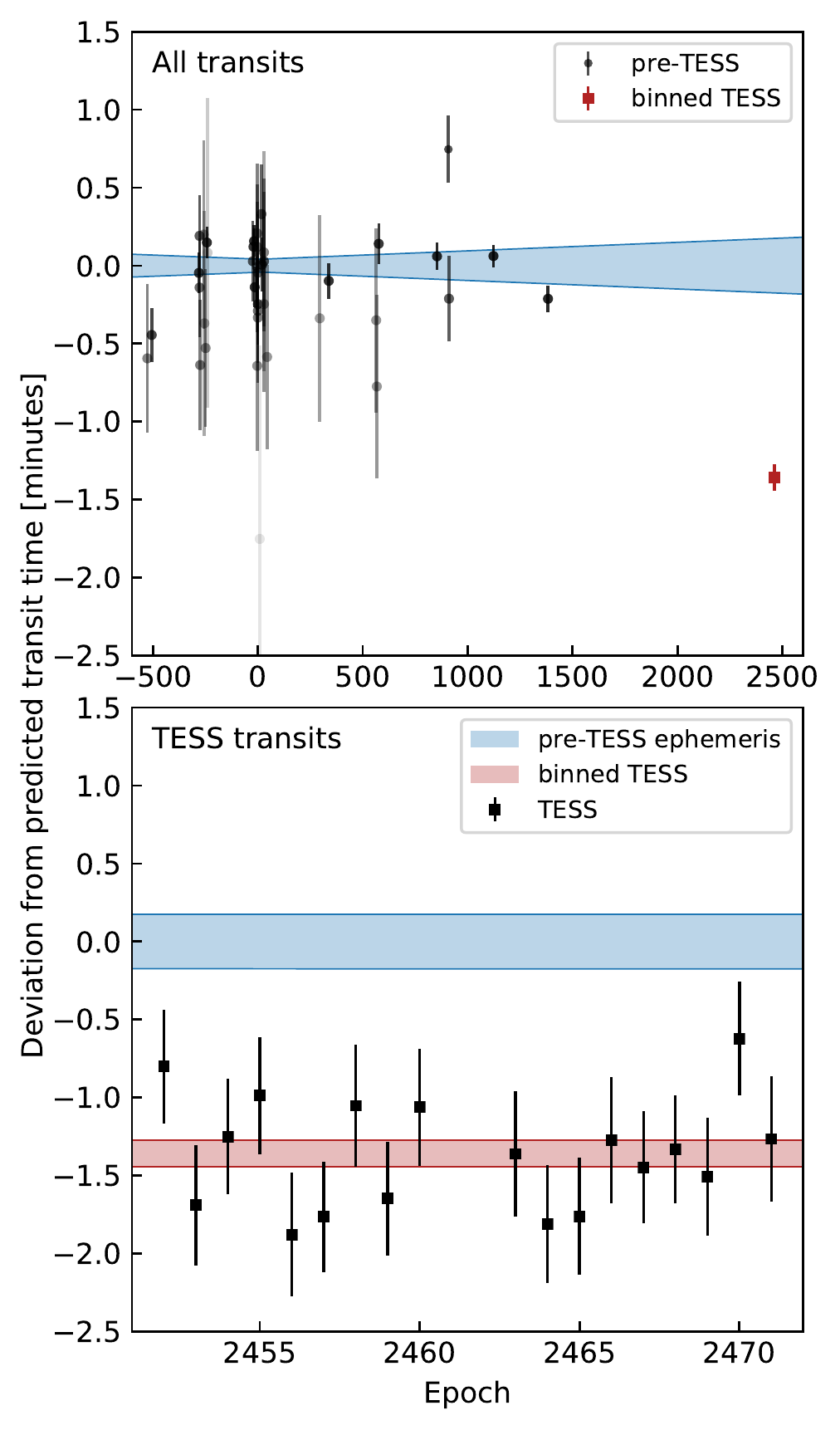}
    \end{center}
    \vspace{-0.6cm}
    \caption{ {\bf TESS saw WASP-4b transit earlier than expected.}
      Both plots show the deviations between observed and calculated
      transit times, where the calculation is based only on
      \deleted{the} pre-TESS data and assumes a constant period.  The
      blue bands depict the $\pm$$1\sigma$ credible interval of the
      predicted times.  {\it Top:} The full timing dataset spans 11
      years. The darkest points correspond to the most precise data.
      \added{The binned TESS point is the weighted average of 18 TESS
      transits.}  {\it Bottom:} Close-up of the TESS observations. The
      red band shows the average deviation of the TESS transits
      \added{($\pm 1\sigma$)}, which arrived $81.6 \pm 11.7\ {\rm
      seconds}$ earlier than predicted.
      \label{fig:arrived_early}
    }
\end{figure}

\begin{figure*}[t]
	\begin{center}
		\leavevmode
		\includegraphics[width=0.9\textwidth]{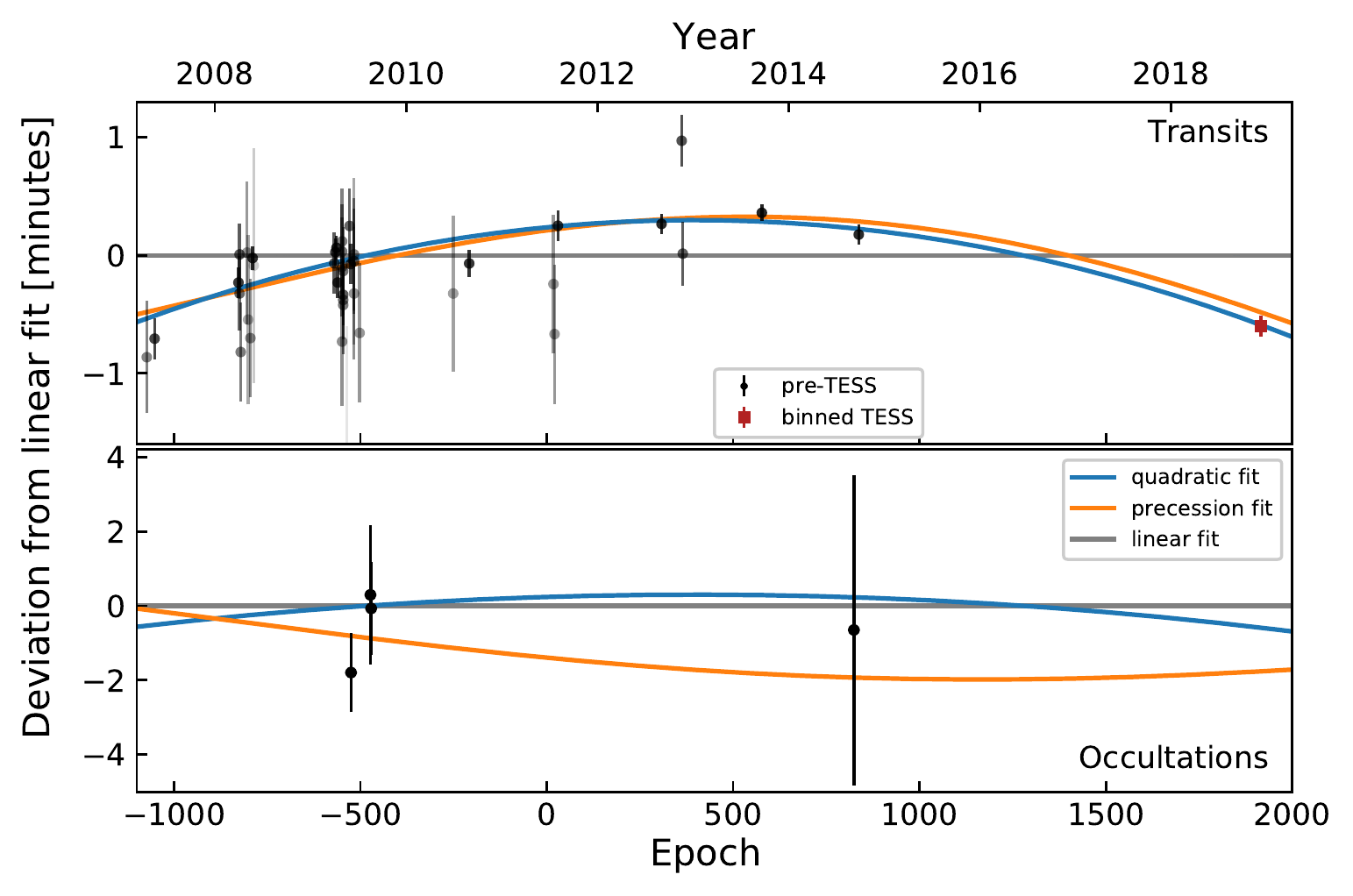}
	\end{center}
	\vspace{-0.7cm}
  \caption{ {\bf Timing residuals and best-fit models for WASP-4b.}
  \replaced{The residuals are the observed times minus the calculated
  times assuming a constant period. The darkest points correspond to
  the most precise data.}{The vertical axis shows the observed times
  minus the calculated times assuming a constant period for transits
  ({\it top}) and occultations ({\it bottom}). In the upper panel,
  darker points correspond to more precise data.} The constant-period
  model (gray line) is a poor description of the data.  Models with a
  decreasing period (blue) or an eccentric, precessing orbit (orange)
  provide	better fits\deleted{ to the data}.  The binned TESS
  point (red square) is the weighted average of 18 TESS transits
  and is for display purposes only.
  The models were fitted to all of the individual transit times.
  \label{fig:times}
	}
\end{figure*}

\subsection{Pre-TESS timing measurements}
\label{subsec:times}

Table~2 gives the transit times we used in our analysis. We included
data from peer-reviewed literature for which the analysis was based on
observations of a single\deleted{, complete} transit, and for which
the midpoint was allowed to be a free parameter. We also required that
the time system be clearly documented. Many of the times were
previously compiled by \citet{hoyer_tramos_2013}. We confirmed that
the times in that paper were in agreement with the original sources
and that barycentric corrections had been performed when needed.

The earliest epoch is from EulerCam on the 1.2-m Euler telescope
\citep{wilson_wasp-4b_2008}.  The second epoch is based on $z$-band
photometry acquired by \citet{gillon_improved_2009} at the VLT 8.2-m
with FORS2.  Subsequent observations were performed by
\citet{winn_transit_2009}, \citet{dragomir_terms_2011},
\citet{sanchis-ojeda_starspots_2011}, \citet{nikolov_wasp-4b_2012},
\citet{hoyer_tramos_2013}, and \citet{ranjan_atmospheric_2014}.
Finally, \citet{huitson_gemini_2017} acquired optical transit spectra
with the 8.1-m Gemini South telescope between 2011 and 2014, one
transit per season.  The per-point standard deviation of their
lightcurves was a few hundred parts per million.  The average
precision in their reported transit times was 5.6~seconds.  Since
these data points carry significant weight in the analysis, we
\replaced{checked}{corresponded with the authors to confirm} that the
timestamps in their data represent mid-exposure times, that the
barycentric correction was performed correctly, and that the time
system of the final results was BJD$_{\rm TDB}$.  These same authors
also used the same instrument and method to analyze other hot
Jupiters, none of which showed a departure from a constant-period
model.  Finally, these authors also measured two transit midpoints
using Spitzer (Baxter et al., in prep); these times agree with the
Gemini South results.

We also compiled the available occultation times, which are given in
Table~3.  The tabulated values have been corrected for the
light-travel time across the diameter of the orbit by subtracting
$2a/c = 22.8$ seconds from the observed time.
\citet{beerer_secondary_2011} observed two occultations of WASP-4b
using warm Spitzer in the 3.6\,$\mu$m and 4.5\,$\mu$m bands.
\citet{caceres_ground-based_2011} detected an occultation from the
ground in the $K_S$ band, and gave a time in HJD, without specifying
the time standard.  We assumed the standard was UTC, and performed the
appropriate corrections to convert to BJD$_{\rm TDB}$.  We verified
that none of our conclusions would be changed if this assumption was
mistaken. Finally, \citet{zhou_secondary_2015} observed an occultation
with the Anglo-Australian Telescope.  They did not report the observed
midpoint, but they did report a result for $e\cos\omega$ based upon
the observed midpoint.  We calculated the implied midpoint using the
formula \citep[{\it e.g.},][]{winn_exoplanet_2010}
\begin{equation}
  t_{\rm occ}(E) =
  t_0 +  P E  +
  \frac{P}{2} \left( 1 + \frac{4}{\pi} e\cos\omega \right),
  \label{eq:occultation_time}
\end{equation}
\added{for $E$ the transit number, $t_0$ the reference epoch, $e$ the
eccentricity, and $\omega$ the argument of pericenter.} In total,
there are four \added{available} occultation times.

\subsection{Analysis}
\label{subsec:analysis}

First, we \replaced{fitted}{performed a weighted least-squares fit of}
a constant-period model (\added{a }``linear ephemeris'') to the
pre-TESS data, and used it to extrapolate to the epochs of the TESS
observations.  The residuals of the best fitting model are shown in
Figure~\ref{fig:arrived_early}.  The transits observed by TESS
occurred earlier than expected.  Because the TESS mission is still in
an early stage, we were concerned about a possible offset in the TESS
timestamps due to an error with the TESS clock or the data processing
pipeline.  Appendix~\ref{sec:verify_tess} describes some tests that
convinced us that a simple offset is unlikely. Assuming that the
observed timing variation is astrophysical, we proceeded by exploring
three models for the timing data in a manner identical to the study
by~\citet{patra_2017}.

The first model assumes a constant orbital period on a circular orbit:
\begin{align}
  t_{\rm tra}(E) &= t_0 + PE,\\
  t_{\rm occ}(E) &= t_0 + \frac{P}{2} + PE,
\end{align}
where $E$ is the epoch number.  We defined the epoch numbers such that
$E=0$ is near the weighted average of the observed times.  This helps
to reduce the covariance between $t_0$ and $P$.

The second model assumes the period is changing at a steady rate:
\begin{align}
  t_{\rm tra}(E) &=
    t_0 + PE +
    \frac{1}{2} \frac{{\rm d}P}{{\rm d}E} E^2, \\
  t_{\rm occ}(E) &=
    t_0 + \frac{P}{2} + PE +
    \frac{1}{2} \frac{{\rm d}P}{{\rm d}E} E^2.
\end{align}
The three free parameters are the reference epoch $t_0$, the period at
the reference epoch, and the period derivative, ${\rm d}P/{\rm d}t =
(1/P) {\rm d}P/{\rm d}E$.

The third model assumes the planet has a slightly eccentric orbit, and
that the line of apsides is rotating \citep{gimenez_revision_1995}:
\begin{align}
  t_{\rm tra}(E) &= 
		t_0 + P_{\rm s}E
    - \frac{e P_{\rm a}}{\pi} \cos\omega,\\
  t_{\rm occ}(E) &= 
    t_0 + \frac{P_{\rm a}}{2} + P_{\rm s}E
    + \frac{e P_{\rm a}}{\pi} \cos\omega,
\end{align}
where $P_{\rm s}$ is the sidereal period, $e$ is the eccentricity,
$P_{\rm a}$ is the anomalistic period, and $\omega$ is the argument of
pericenter.  In this model the angular velocity of the line of apsides
${\rm d}\omega/{\rm d}E$ is constant,
\begin{equation}
  \omega(E) = \omega_0 + \frac{{\rm d}\omega}{{\rm d}E} E,
\end{equation}
and the sidereal and anomalistic periods are connected through the
equation
\begin{equation}
  P_{\rm s} = P_{\rm a} \left(
    1 - \frac{1}{2\pi}\frac{{\rm d}\omega}{{\rm d}E}
    \right).
\end{equation}
\added{The sidereal period is the duration required to return to the
same orientation with respect to the stars; the slightly longer
anomalistic period is the duration required to reach a fixed longitude
with respect to the rotating line of apsides.} The five free
parameters of this model are $(t_0, P_{\rm s}, e, \omega_0, {\rm
d}\omega/{\rm d}E)$, denoting the reference epoch, the sidereal
period, the eccentricity, the argument of pericenter at the reference
epoch, and the angular velocity of the line of apsides.

\added{We fitted each model by assuming a Gaussian likelihood and
sampling over the posterior probability distributions. The prior for
the quadratic model allowed the period derivative to have any sign.
We considered two possible priors for the precession model: the first
is a wide prior that allows non-physical values of the planetary
Love number (Equation~\ref{eq:love_number}). The second prior requires
that the planetary Love number is less than that of a sphere of
constant density.  }

Figure~\ref{fig:times} shows the residuals with respect to the
constant-period model.  The best-fitting constant-period model has
$\chi^2 = 174$ and 61 degrees of freedom.  The best-fitting
\replaced{decreasing-period}{quadratic} model has $\chi^2 = 62.6$ and
60 degrees of freedom.  The best-fitting precession model \added{under
the wide prior} has $\chi^2 = 64.3$ and 58
degrees of freedom.\added{ Under the physical prior, the best-fit is
slightly worse, with $\chi^2=64.8$}. The precession and quadratic
models both provide much better fits than the constant-period model.
The difference in $\chi^2$ between the linear and quadratic models
corresponds to $p \approx 10^{-26}$.

The \replaced{decreasing-period}{quadratic} model provides a
marginally better fit to the data than the precession model.  It is
favored by $\Delta \chi^2 = 1.7$, and has two fewer free parameters.
A useful heuristic for model comparison is the Bayesian Information
Criterion,
\begin{equation}
  {\rm BIC} = \chi^2 + k\log n,
\end{equation}
where $k$ is the number of free parameters, and $n$ is the number of
data points. In this case, $n=62$.  The difference in the BIC between
the precession and decay models is $\Delta {\rm BIC} = {\rm BIC}_{\rm
prec} - {\rm BIC}_{\rm quad} = 10$\added{, assuming a wide prior for
the precession model}. This corresponds to a Bayes factor of
$1.1\times 10^{4}$.  Likewise, the Akaike Information Criterion favors
the constant-period-derivative model by $\Delta {\rm AIC} = 5.8$.
Differences of this magnitude are traditionally deemed ``strong
evidence'' that one model is a better description of the data than the
other \citep{kass_bayes_1995}, although we prefer to reserve judgment
until more data can be obtained.

In the \replaced{decreasing-period}{quadratic} model, the period
derivative is
\begin{equation}
\dot{P}
  = - (4.00 \pm 0.38)\times 10^{-10}
  = - 12.6 \pm 1.2~{\rm ms}\,{\rm yr}^{-1}.
  \label{eq:dP_dt_obs}
\end{equation}
For comparison, the best-fitting period derivative of WASP-12b is
$\dot{P} = -29 \pm 3\,{\rm ms}\,{\rm yr}^{-1}$
\citep{maciejewski_departure_2016,patra_2017}.  If both planets are
truly falling onto their stars, then WASP-4b is falling at about half
the rate of WASP-12b.

In the precession model, the best-fit eccentricity is
\begin{equation}
  e = (1.92^{+ 1.93}_{- 0.76})\times10^{-3}
  \label{eq:e_obs}
\end{equation}
The longitude of periastron advances by $\dot{\omega} =
13.6^{+4.7}_{-3.6}\ {\rm degrees}\,{\rm yr}^{-1}$, and the precession
period is $27^{+10}_{-7}$ years.  All of the best-fitting parameters
(the medians of the posterior distributions) and the 68\% credible
intervals are reported in Table~4.

\explain{The following subsection has been added since the original
submission, per the referee's suggestions.}
\subsection{Possible systematic errors}

To assess the overall robustness of our results, we considered a few
possible systematic effects in the timing dataset.  

\paragraph{Suspect pre-TESS light curves}
Some of the pre-TESS light curves have incomplete phase coverage: a
handful of the transit times in Table~2 are from lightcurves with
gaps.  A separate issue is the effect of spot-crossing anomalies on
transit timings.  To address these concerns, we repeated the
model-fitting described above, but omitted epochs -827, -804, -537,
and -208 because of gaps in their coverage.  We also omitted epochs
-526 and -561 because of visible spot anomalies during the transits.
(All epoch numbers are as in Table~2.)  The resulting best-fit transit
timing model parameters were all within $1\sigma$ of the values quoted
in Table~4.  The uncertainties, goodness-of-fit statistics, and model
comparison statistics did not appreciably change.

\paragraph{Spot-crossing events in TESS data}
To explore the effect of possible spot-crossing events on the TESS
transit time measurements, we performed a separate test.  We injected
triangular spot-anomalies with amplitude $0.03\%$ and duration 30
minutes at random phases into each transit.  The amplitude was chosen
to be larger than the spot-crossing anomalies observed by
\citet{southworth_high-precision_2009} and
\citet{sanchis-ojeda_starspots_2011}, and the duration was chosen to
be comparable to those of previously observed events.  Spots of these
amplitudes resemble the possible anomalies present in transits
``1360.54'' and ``1372.58'' of Figure \ref{fig:lightcurves}.

With spots injected, we repeated our measurement of the transit times.
On average, the measured transit times did not change after injecting
spots, because the flux deviations are equally likely to occur in the
first and second halves of the transit.  For individual transits,
there were no cases for which the timing deviation was larger than one
minute.  The largest shifts occur when the spot anomaly occurs during
transit ingress or egress, in which case the measured mid-time is
shifted either late or early by between 30 and 50 seconds.

The TESS observations could therefore be skewed
early if there were spot-crossing events during {\it every} egress.  Two
arguments rule out this possibility.  (1) The lightcurve residuals do
not show evidence for these events.  (2) The stellar rotation period
is between 20 and 40 days, and the sky-projected stellar obliquity is
less than 10 degrees
\citep{triaud_spin-orbit_2010,sanchis-ojeda_starspots_2011,hoyer_tramos_2013}.
Since the planet orbits every 1.3 days, requiring that spot anomalies
always occur during egress would be equivalent to requiring a stellar
spot distribution that is exquisitely (and thus implausibly)
distributed to match the planet egress times.

\paragraph{Detrending choices in pre-TESS data}
There is a final concern that is difficult to address.  We collected
the mid-transit time values derived by different authors, who used
heterogeneous methods to fit and detrend their lightcurves.  We have
also assumed that these authors have correctly documented the time
systems in which the data are reported.  Further, though many choices
in transit-fitting ({\it e.g.}, parametrization of limb-darkening and
eccentricity) do not affect transit mid-time measurements, different
detrending approaches can asymmetrically warp transits and shift
mid-transit times.  The magnitude of this systematic effect is hard to
quantify, but the situation is fairly clear from
Figure~\ref{fig:times}.  Many independent authors provided transit
measurements shortly after WASP-4b's discovery, and the data are
consistent with each other. \citet{huitson_gemini_2017} provided the
most important data from epochs 0-1000. If their data were
systematically affected by detrending choices or time-system confusion
at the level of several times their reported uncertainties, then it
possible that the orbital period is constant despite the evidence in
the TESS data.  For this reason, we paid careful attention to the
\citet{huitson_gemini_2017} data set, and corresponded with the
authors to confirm that their results are not affected by systematic
effects of the required amplitude.

None of the concerns mentioned in this subsection seem likely to
explain the observed timing variations. We proceed by considering
possible astrophysical explanations.

%%%%%%%%%%%%%%%%%%%%%%%%%%%%%%%%%%%%%%%%%%
\section{Interpretation}
\label{sec:implications}

\begin{figure*}[t]
  \begin{center}
    \includegraphics[height=0.88\textheight]{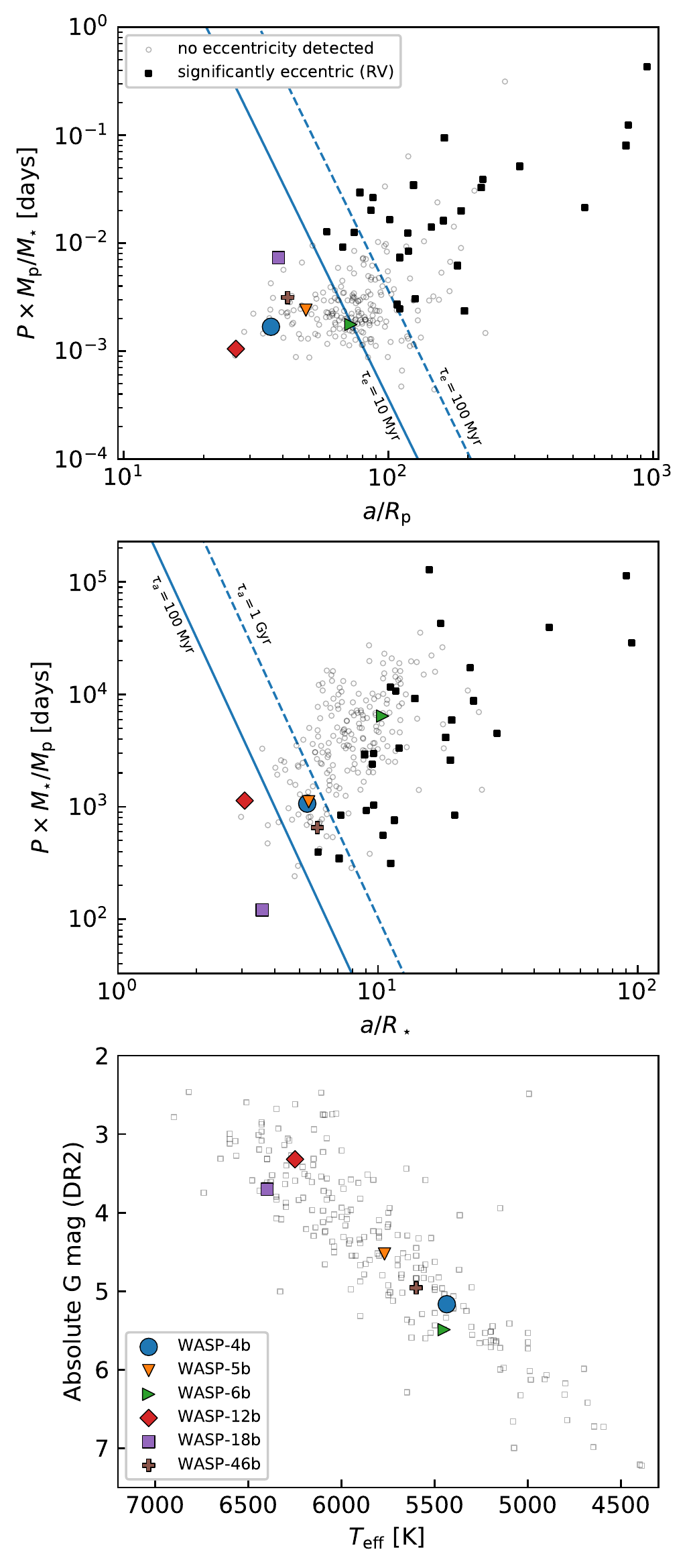}
  \end{center}
  \vspace{-0.6cm}
  \caption{
    {\bf WASP-4b in the context of other hot Jupiters.}
    Most of the data in these plots are from \citet{bonomo_gaps_2017},
    who measured the eccentricities using radial velocities.  The
    colored symbols highlight WASP-4, WASP-12 (which also shows
    evidence for a decreasing period), and the\deleted{ collection of}
    hot Jupiters analyzed in Appendix A.
    {\bf \it Top:}
    Key parameters relevant to eccentricity damping.  Solid squares
    represent planets with a securely detected nonzero eccentricity.
    \added{Circles represent planets whose orbits are consistent with
    circular.} Lines of constant damping timescale are drawn based on
    Equation~\ref{eq:de_dt} and assuming $Q_{\rm p}' = 10^5$.  WASP-4b
    has one of the shortest eccentricity damping times.
    {\bf \it Middle:} 
    Key parameters relevant to orbital decay.  Lines of constant decay
    timescale are drawn based on Equation~\ref{eq:da_dt}, assuming
    $Q_\star' = 10^7$.  WASP-4b has a relatively short orbital decay
    timescale, although it is not as extreme a case in this regard as
    it is for eccentricity damping.
    {\bf \it Bottom:}
    A Hertzsprung-Russell diagram for hot Jupiter hosts. WASP-4
    appears to be on the main sequence.
    \label{fig:context}
  }
\end{figure*}

\subsection{Orbital decay}

If the timing variation is \deleted{being caused }caused entirely by
orbital decay, then \replaced{the}{the best-fit model parameters yield a}
characteristic\replaced{ timescale of the decay is}{ decay timescale of}
\begin{equation}
  \frac{P}{ {\rm d}P/{\rm d}t } = 9.2 \, {\rm Myr}.
\end{equation}
For comparison, the corresponding time for WASP-12b is 3.2\,Myr
\citep{patra_2017}.

If WASP-4 really is undergoing rapid orbital decay, then how many of
the other known hot Jupiters should have orbits that are decaying at
detectable (or nearly detectable) rates?  Figure~\ref{fig:context}
compares some key properties of WASP-4 with those of a larger ensemble
of hot Jupiters.  The middle panel displays two parameters that
strongly affect the expected orbital decay timescale, $P
M_\star/M_{\rm p}$ and $a/R_\star$.  WASP-4 has one of the shortest
theoretical timescales for orbital decay.  \replaced{There
are}{Figure~\ref{fig:context} shows} about 20 hot Jupiters (including
WASP-12) for which the theoretical timescale is shorter.  In almost
all of those cases, though, the planet was discovered more recently
than WASP-4 and a decade-long baseline of observations is not yet
available.  A separate consideration not shown in
Figure~\ref{fig:context} is that the hot Jupiter host stars have a
variety of different structures, from being fully convective to nearly
fully radiative, which may lead to widely divergent tidal dissipation
timescales.

In the simple ``constant phase lag'' model for tidal interaction
\citep{zahn_tidal_1977}, the rate of dissipation can be parametrized
by a modified\footnote{For stars, $k_\star \sim \mathcal{O}(10^{-2})$,
so it is important to explicitly distinguish $Q_\star'$ from $Q_\star$
\citep[{\it e.g.},][]{schwarzschild_structure_1958}.} quality factor,
$Q_\star' = 3 Q_\star / (2k_\star)$.  Here, $Q_\star$ is the ratio
between the energy stored in the equilibrium deformation of the star
and the energy lost to heat per tidal period \citep[{\it
e.g.},][]{goldreich_q_1966}.  A larger $Q_\star$ implies less
efficient tidal dissipiation. The dimensionless number $k_\star$ is
the stellar Love number, which is smaller when the star's density
distribution is more centrally concentrated.  In this model, once the
planet's spin and orbit are synchronized, then the semi-major axis and
eccentricity evolve as \citep[Appendix B of][]{metzger_optical_2012}
\begin{align}
  \frac{1}{\tau_{\rm e}} &=
  \frac{|\dot{e}|}{e} =
    \frac{63 \pi } {2 Q_{\rm p}' }
    \left( \frac{R_{\rm p}}{a} \right)^5
    \left( \frac{M_\star}{M_{\rm p}} \right)
    \left( \frac{1}{P} \right)
  \label{eq:de_dt}
  \\
  \frac{1}{\tau_{\rm a}} &=
  \frac{|\dot{a}|}{a} =
    \frac{9 \pi } {Q_\star' }
    \left( \frac{R_\star}{a} \right)^5
    \left( \frac{M_{\rm p}}{M_\star} \right)
    \left( \frac{1}{P} \right).
  \label{eq:da_dt}
\end{align}
The orbital period evolves as
\begin{equation}
\label{eq:dP_dt}
  \dot{P} = -\frac{27\pi}{2 Q_\star'}
            \left(\frac{M_{\rm p}}{M_\star}\right)
            \left(\frac{R_\star}{a}\right)^5.
\end{equation}
The modified quality factor of WASP-4 corresponding to the observed
value of $\dot{P}$ is
\begin{equation}
  Q_\star' = (2.9 \pm 0.3)\times10^4.
\end{equation}
This is about an order of magnitude lower than the value that was
inferred for WASP-12b.  It is also smaller than most theoreticians
would have expected.  The $Q_{\rm Jup}'$ value of Jupiter is estimated
to be $\approx$$1.4 \times 10^5$, based on the observed motions of the
Galilean moons \citep{lainey_strong_2009}.  For stars, studies of the
binary eccentricity distribution have been interpreted with tidal
models, giving $Q_\star' \approx 10^5 - 10^7$ \citep[{\it
e.g.},][]{meibom_robust_2005,belczynski_compact_2008,
geller_direct_2013,milliman_wiyn_2014}.  Population studies of hot
Jupiter systems have also been undertaken, generally finding $Q_\star'
\approx 10^5 - 10^8$ using different models
\citep{jackson_observational_2009,hansen_calibration_2010,penev_constraining_2012,penev_empirical_2018,cameron_hierarchical_2018}.
For instance, motivated by the rapid rotation of some hot Jupiter
hosts
\citep{pont_empirical_2009,ciceri_hats-15b_2016,penev_hats-18b_2016},
\citet{penev_empirical_2018} modeled the evolution of hot Jupiter
systems under the influence of a magnetized wind and a constant
phase-lag tide.  For WASP-4, their method gave $Q_\star' \approx
(1.2^{+1.0}_{-0.5})\times10^7$, which would correspond to $\dot{P}
\approx -30$ microseconds per year.  This strongly disagrees with the
the period change that we have observed.

\citet{essick_orbital_2016} studied the problem of the orbital decay
of hot Jupiters using a theory in which gravity modes are excited at
the base of the stellar convective zone, propagate inward through the
radiative core and break near the stellar core, leading to energy
dissipation.  They predicted the stellar quality factors in hot
Jupiter systems to vary from $Q_\star' \approx 10^5 - 10^6$.  From
their Equation 26, the prediction for WASP-4 is $Q_\star' =
7\times10^5$, which is an order of of magnitude larger than implied by
the observed period change.

The applicability of the \citet{essick_orbital_2016} model depends on
the evolutionary state of the star. \citet{weinberg_tidal_2017} showed
that more rapid dissipation --- enough to account for the period
change of WASP-12b --- could exist in stars that have begun evolving
into red giants.  The bottom panel of Figure~\ref{fig:context} shows a
Hertzsprung-Russell diagram of hot Jupiters hosts, including WASP-4.
On the y-axis is $G=g-\mu$, for $g$ the apparent {\it Gaia}-band magnitude,
and $\mu$ the distance modulus reported by
\citet{gaia_collaboration_gaia_2018}.  The x-axis is the effective
temperature from \citet{bonomo_gaps_2017}, which for WASP-4 agrees
within $1\sigma$ of that from Table~1.  Inspecting the HR diagram,
WASP-4 shows little evidence of being evolved, in agreement with our
analysis from \S~\ref{sec:system_parameters}.
%%%%%%%%%%%%%%%%%%%%%%%%%%%%%%%%%%%%%%%%%%
% By a similar argument though, WASP-12 would not clearly stand out as a
% candidate subgiant, and \citet{weinberg_tidal_2017} suggest it may in
% fact be a subgiant (though \citealt{bailey_understanding_2019} do not
% find evidence that supports this suggestion).  A more thorough
% isochronal analysis would be of interest, particularly if the transit
% and occultation times continue to vary.  Such an analysis is beyond
% the scope of this study.
%%%%%%%%%%%%%%%%%%%%%%%%%%%%%%%%%%%%%%%%%%

To summarize, if the observed period change is caused entirely by
tidal orbital decay, then the constant-phase-lag tidal model implies a
stellar tidal dissipation rate that is higher than expected by at
least an order of magnitude.  It might be possible that we are
observing at a special time, shortly after the planet's inward
migration, or when the planet is near resonance with a stellar
oscillation mode.  Tidal dissipation rates might also be increased if
the star is just turning off the main sequence.  Another hypothesis,
recently advanced by \citet{millholland_obliquity_2018} for the case
of WASP-12b, is that an exterior planet could be trapping WASP-4b's
spin vector in a high-obliquity state, leading to rapid dissipation
through planetary obliquity tides.

\subsection{Apsidal precession}
\label{sec:apsidal_precession}

\explain{The result given in Equation~\ref{eq:love_number} has been
corrected from the submitted version. The previously stated result, k2
= $1.5^{+2.1}_{-1.1}$ had a
typo that led to printing of 84th and 16th percentiles, instead of the
difference of these percentiles from the median. 
The uncertainties also previously did not propagate uncertainty in the
planet radius and semimajor axis, which enter at the fifth power and
are now included. 
Finally, we now also specify the assumed priors, as they add a
nuance in the interpretation.   }

If instead the observed timing variation is just a small portion of an
apsidal precession cycle, then the orbital eccentricity is a few times
$10^{-3}$, and the full precession period is about 27 years.
\citet{ragozzine_probing_2009} calculated apsidal precession periods
for hot Jupiters, finding them to range between about 10 and 100
years. They highlighted that for many hot Jupiters, including WASP-4b,
the theoretical precession rate is dominated by the non-Keplerian
force due to the planet's tidal bulge.  Precession from general
relativity, the planet's rotational bulge, and the star's rotational
and tidal bulges contribute at the 10\% level at most.  Thus, a
measurement of the precession rate can be used to determine the
planet's Love number.  From their Equation 14, the implied Love number
for WASP-4b is
\begin{align}
  k_{2,{\rm p}} &= 1.59^{+0.70}_{-0.47}
    \quad{\rm (wide\ prior),} \label{eq:love_number} \\
  k_{2,{\rm p}} &=  1.20^{+0.20}_{-0.26}
    \quad(k_{2,{\rm p}} \sim \mathcal{U}[0.015, 1.5]),
\end{align}
where the uncertainties in the semimajor axis and planet radius have been
propagated through our Markov chains, and $\mathcal{U}$ denotes the uniform 
distribution.  For comparison, the Love number
of Jupiter is about 0.55 \citep{wahl_tidal_2016,ni_empirical_2018},
and a uniform density sphere has $k_2 = 1.5$. The uncertainty in
$k_{2,{\rm p}}$ for WASP-4b is large because the eccentricity,
reference time, and ${\rm d}\omega/{\rm d}E$ have strongly correlated
errors, and the \added{four available}\deleted{measured} occultation
times only weakly constrain these parameters
(Figure~\ref{fig:future}).

\added{ The results under both priors are larger than the Love number
of Jupiter, at weak statistical signifiance.  Imposing the
physically-motivated prior truncates the allowed values of $k_{2,{\rm
p}}$, and drives the precession period towards larger values.
Different priors therefore change the predicted evolution of the
orbit.  While the precession model matches the orbital decay model for
the next two decades in transits (Figure~\ref{fig:future}), in
occultations the precession model can deviate from the orbital decay
model, particularly if one requires the planetary Love number to be
smaller. This suggests that promptly obtaining occultation time
measurements for the system could help rule out the apsidal precession
model.  For the time being, the data are insufficient to make a
stronger judgement.  }

\added{Regardless of the detailed model assumptions, the}
\deleted{The} main \added{physical }problem with the apsidal precession
hypothesis is to explain why the
eccentricity would be as large as $\sim$$10^{-3}$ despite rapid tidal
circularization.  For WASP-4, Equation~\ref{eq:de_dt} gives $\tau_e =
0.29 (Q_{\rm p}'/10^5) \, {\rm Myr}$.  The star is several billion
years old, so unless the planet arrived very recently, any initial
eccentricity should have been lowered well below $\sim$$10^{-3}$.  The
top panel of Figure~\ref{fig:context} compares the expected
eccentricity damping time of WASP-4b with that of other transiting
giant planets.  WASP-4b has one of the shortest known eccentricity
damping times.

\paragraph{Neighboring companion}
One way to maintain a significant eccentricity is through the
gravitational perturbations from another planet.
\citet{mardling_long-term_2007} considered the long-term tidal
evolution of hot Jupiters with companions.  The companion in their
model is coplanar, and can have a mass down to an Earth-mass; the main
requirement is that both the hot Jupiter and the outer companion start
on eccentric orbits.  They found that although the early phases of the
two-planet eccentricity evolution occur quickly, the final phase of
the joint eccentricity evolution towards circularity would occur on
timescales several orders of magnitude longer than the circularization
time of an isolated hot Jupiter (see their Figures~4 and 5).
 
A separate way a neighboring companion could excite the hot Jupiter's
eccentricity is through the
Kozai-Lidov mechanism \citep{lidov_evolution_1962,kozai_secular_1962}.
In this case, the orbital plane of the outer companion, ``c'', would
need to be inclined relative to that of the hot Jupiter, ``b'', by at
least $\sin^{-1} \sqrt{2/5} \approx 39^\circ$.
For the Kozai-Lidov mechanism to operate at maximum efficiency, we
need \citep[][Equation 20]{bailey_understanding_2019}
\begin{equation}
  M_{\rm c} > 7.5\,M_\oplus
  \times \left( \frac{a_{\rm c}}{a_{\rm b}} \right)^{3/2},
  \label{eq:kozai_bound}
\end{equation}
\replaced{where $M_{\rm b,c}$ and $a_{\rm b,c}$ are the mass and semi-major 
axis of WASP-4b and the hypothetical WASP-4c, and owing}{where $a_{\rm b}$ is 
the semi-major axis of WASP-4b, $M_{\rm c}$ is the mass of the hypothetical 
WASP-4c, and $a_{\rm c}$ is WASP-4c's semi-major axis. Owing}
to our imprecise
measurement\added{, in Equation~\ref{eq:kozai_bound}} we have assumed 
WASP-4b's Love number is $k_{2,{\rm
b}}\approx 0.6$\added{, similar to Jupiter}.  For the RV signal of the 
companion to remain
undetected, it would need to be in the residual $({\rm O-C})_{\rm RV}
= 15.2\,{\rm m s}^{-1}$ reported by \citet{triaud_spin-orbit_2010}.
Again following \citet{bailey_understanding_2019}, this implies
\begin{align}
  M_{\rm c} &<
  ({\rm O-C})_{\rm RV}
  \left( \frac{M_\star a_b}{G} \right)^{1/2}
  \left( \frac{a_c}{a_b} \right)^{1/2}
  f^{-1/2}
  \nonumber
  \\
  M_{\rm c} &< 
  23.7\,M_\oplus
  \times 
  \left( \frac{a_c}{a_b} \right)^{1/2}
  f^{-1/2},
  \label{eq:rv_bound}
\end{align}
for $f(e_{\rm c},\omega_{\rm c}, i_{\rm c}) \propto \sin^2 i_{\rm c}$
a geometric prefactor that depends on the argument of periastron
$\omega_{\rm c}$ and inclination $i_{\rm c}$ of the exterior companion
(\citealt{bailey_understanding_2019} Equation 23).  Since WASP-4b is
transiting, $f$ can be arbitrarily small, and both of the preceding
limits can be satisfied. 

\paragraph{Fluctuations in the gravitational potential from convection}
An independent mechanism to pump the eccentricity invokes the
gravitational fluctuations from stellar convection \citep[][Section
7]{phinney_pulsars_1992}.  From equation 7.33 of that work, the
mean-squared eccentricity of the orbit is
\begin{equation}
  \langle e^2 \rangle =
  \frac{ 2 \langle E_{\rm e} \rangle }{\mu n^2 a^2}
  = 6.8\times10^{-5}
  \frac{(L^2 R_{\rm conv}^2 M_{\rm conv}^2)^{1/3}}{\mu n^2 a^2},
\end{equation}
where $L$ is the stellar luminosity, $R_{\rm conv}$ and $M_{\rm conv}$
are the width and mass of the convective region, $\mu$ is the reduced
mass, $n$ is the orbital frequency, and $a$ is the semi-major axis.
\added{For the luminosity, reduced mass, and semi-major axis, we used
values from Table~1, combined with the Stefan-Boltzmann law and
standard definitions. To estimate the width and mass of the convective
region we ran the MESA code for a star with mass and metallicity
matched to WASP-4, and the input physics detailed in the MIST
isochrones project
\citep{paxton_modules_2011,paxton_modules_2013,paxton_modules_2015,dotter_mesa_2016,choi_mesa_2016}.
We identified the tachocline boundary using the mixing types specified
in the resulting radial profiles, and found $R_{\rm conv} \approx 0.33
R_\odot$, and $M_{\rm conv} \approx 9\times10^{-4} M_\odot$. } For
WASP-4, this implies $\langle e^2 \rangle^{1/2} \lesssim 10^{-5}$.
Hence, this mechanism does not seem capable of producing the required
eccentricity of $\sim$$10^{-3}$.

\explain{The subsequent subsection adds a description of our RV
analysis, which we did not include in the original submission. Some of
the material was taken from what was previously the "Other possible
explanations" section.}
\subsection{Timing variation due to line of sight acceleration}

\begin{figure}[t]
	\begin{center}
		\leavevmode
		\includegraphics[width=0.48\textwidth]{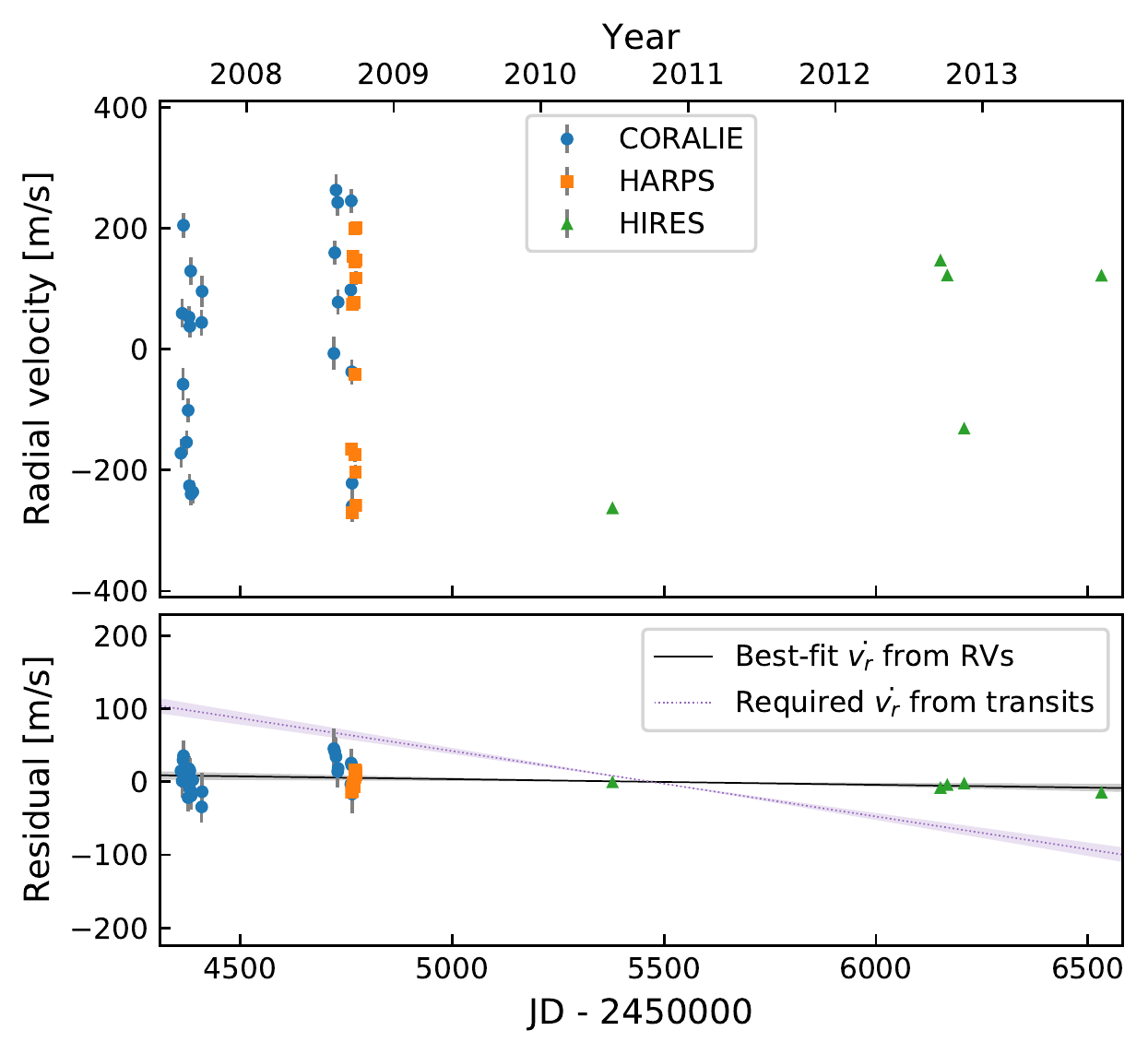}
	\end{center}
	\vspace{-0.7cm}
	\caption{
    {\bf Radial velocities of WASP-4 (top), and residuals from the
    best-fit Keplerian model (bottom).} The lower panel shows the
    best-fit linear trend inferred from the RV data (black line,
    $1\sigma$ errors in gray), and the trend that would be needed to
    produce the period decrease seen in transits (purple dotted line).
    Since both the RV and transit timing datasets are sparse after
    2013, a distant massive companion on an eccentric orbit might
    still explain the observations.
	\label{fig:rvs}
  \vspace{-0.3cm}
	}
\end{figure}

An acceleration of the center of mass of the system towards our line
of sight could cause a decrease in the apparent orbital period.  The
period derivative would be
\begin{equation}
	\dot{P} = \frac{\dot{v}_{\rm r} P}{c},
\end{equation}
where $\dot{v}_{\rm r}$ is the time derivative of the radial velocity.

\citet{knutson_friends_2014}, combining radial-velocity data from
their own program with those of \citet{wilson_wasp-4b_2008},
\citet{pont_determining_2011}, and \citet{husnoo_observational_2012},
found evidence for long-term trend in WASP-4, with low statistical
significance:
\begin{equation}
  \dot{v}_{\rm r} =
     -0.0099^{+0.0052}_{-0.0054}
     \,{\rm m}\,{\rm s}^{-1}\,{\rm day}^{-1}.
\end{equation}
This acceleration translates to an expected $\dot{P} = -(4.4 \pm 2.4)
\times10^{-11}$. The period decrease from the observed RV trend is an
order of magnitude smaller than the observed $\dot{P}$ from transit
times, $(4.0\pm 0.4)\times10^{-10}$ (Table~4).

Nonetheless, it is intriguing that the host star shows a weakly
significant acceleration, and with the correct sign needed to explain
the transit timing variations.  Given the potential importance of the
radial velocities for interpreting this system, we performed an
independent analysis, as follows.  

First, we collected the usable RV measurements from CORALIE, HARPS,
and HIRES.  We included the CORALIE measurements from
\citet{wilson_wasp-4b_2008} and \citet{triaud_spin-orbit_2010}, using
the homogeneous radial velocities calculated by the latter authors. We
included the HARPS values reported by \citet{pont_determining_2011},
which are identical to those from \citet{husnoo_observational_2012}.
We omitted the HARPS data points taken over three nights by
\citet{triaud_spin-orbit_2010} for Rossiter-McLaughlin observations
because they were calculated using a different pipeline than the
longer-baseline \citeauthor{pont_determining_2011} HARPS measurements,
and necessary inclusion of an extra offset term would nullify their
statistical value.  Finally, we included the five HIRES measurements
taken over many years by \citet{knutson_friends_2014}.  

We then fitted a single Keplerian orbit, plus instrument offsets,
jitters, and a long-term trend
\citep[][\texttt{radvel}]{fulton_radvel_2018}.  We set Gaussian priors
on the period and time of inferior conjunction using the values from
Table~4, and fixed the eccentricity to zero, consistent with results
from \citet{beerer_secondary_2011}, \citet{knutson_friends_2014} and
\citet{bonomo_gaps_2017}.  The remaining free parameters were the
velocity semi-amplitude, the instrument zero-points, the instrument
jitters (an additive white noise term for each instrument), linear
($\dot{v_{\rm r}}$), and optionally second-order ($\ddot{v_{\rm r}}$)
acceleration terms.

We found that the best-fitting model with both linear and quadratic
radial velocity terms was marginally preferred (by $\Delta
\mathrm{BIC} = 5.8$) over the best-fitting model with only a linear
term.  Regardless, for consistency with \citet{knutson_friends_2014},
who fixed the quadratic component of the long-term trend to zero, in
Figure~\ref{fig:rvs} we show best-fitting models for the linear-trend
case.  The best-fit value for the line of sight acceleration,
\begin{equation}
  \dot{v}_{\rm r} =
     -0.0077^{+0.0052}_{-0.0047}
     \,{\rm m}\,{\rm s}^{-1}\,{\rm day}^{-1},
\end{equation}
is within $1\sigma$ of the value reported by
\citet{knutson_friends_2014}. (Note that the CORALIE data used in our
and their analyses differ, as we included additional measurements
reported by \citealt{triaud_spin-orbit_2010}).  The implied period
derivative is still therefore about an order of magnitude smaller than
our observed $\dot{P}$ from transit timing.

To summarize, only about one tenth of the observed period decrease can
be explained through a constant acceleration of the WASP-4 system's
center of mass.  However, given the limited amount and uneven time
coverage of the existing radial-velocity data (Figure~\ref{fig:rvs}),
it remains possible that the center of mass has a more complex motion,
perhaps due to a companion on an eccentric orbit \citep[{\it e.g.},
WASP-53 or WASP-81,][]{triaud_peculiar_2017}.  It would be useful to
gather more radial-velocity data to confirm or refute this
possibility.

\subsection{Applegate effect}

A separate candidate explanation for the timing deviations is the
\citet{applegate_mechanism_1992} effect.  Some eclipsing binaries
exhibit period modulations with amplitudes of $\lesssim 0.05\,{\rm
days}$ over timescales of decades \citep[{\it
e.g.},][]{soderhjelm_geometry_1980,hall_relation_1989}.  The Applegate
mechanism explains these modulations by positing that the internal
structure of a magnetically active star changes shape via cyclic
exchange of angular momentum between the inner and outer zones of the
star.  This model could also apply to a hot Jupiter orbiting a star
with a convective zone.  The changing gravitational quadrupole of the
star would cause the orbit of the planet to precess on the timescale
of the stellar activity cycle.  An essential difference between this
process and apsidal precession is that Applegate timing variations
need not be strictly periodic \citep[{\it
e.g.},][Figure~12]{soderhjelm_geometry_1980}. This mechanism would
also produce transit and occultation timing deviations of the same
sign, while for apsidal precession they would have opposite signs.
For WASP-4, \citet{watson_orbital_2010} estimated that the Applegate
effect could produce timing deviations of up to 15 seconds, depending
on the modulation period of the stellar dynamo.  If this analysis is
accurate, then the Applegate mechanism cannot explain the majority our
observed $82$ second variation.

\explain{The following subsection had material moved into the ``Timing
variation due to line of sight acceleration'' subsection. Figure 7 was
updated to clarify the Love number prior for the precession model.}
\subsection{Other possible explanations}

\begin{figure*}[!ht]
	\begin{center}
		\leavevmode
		\includegraphics[width=0.9\textwidth]{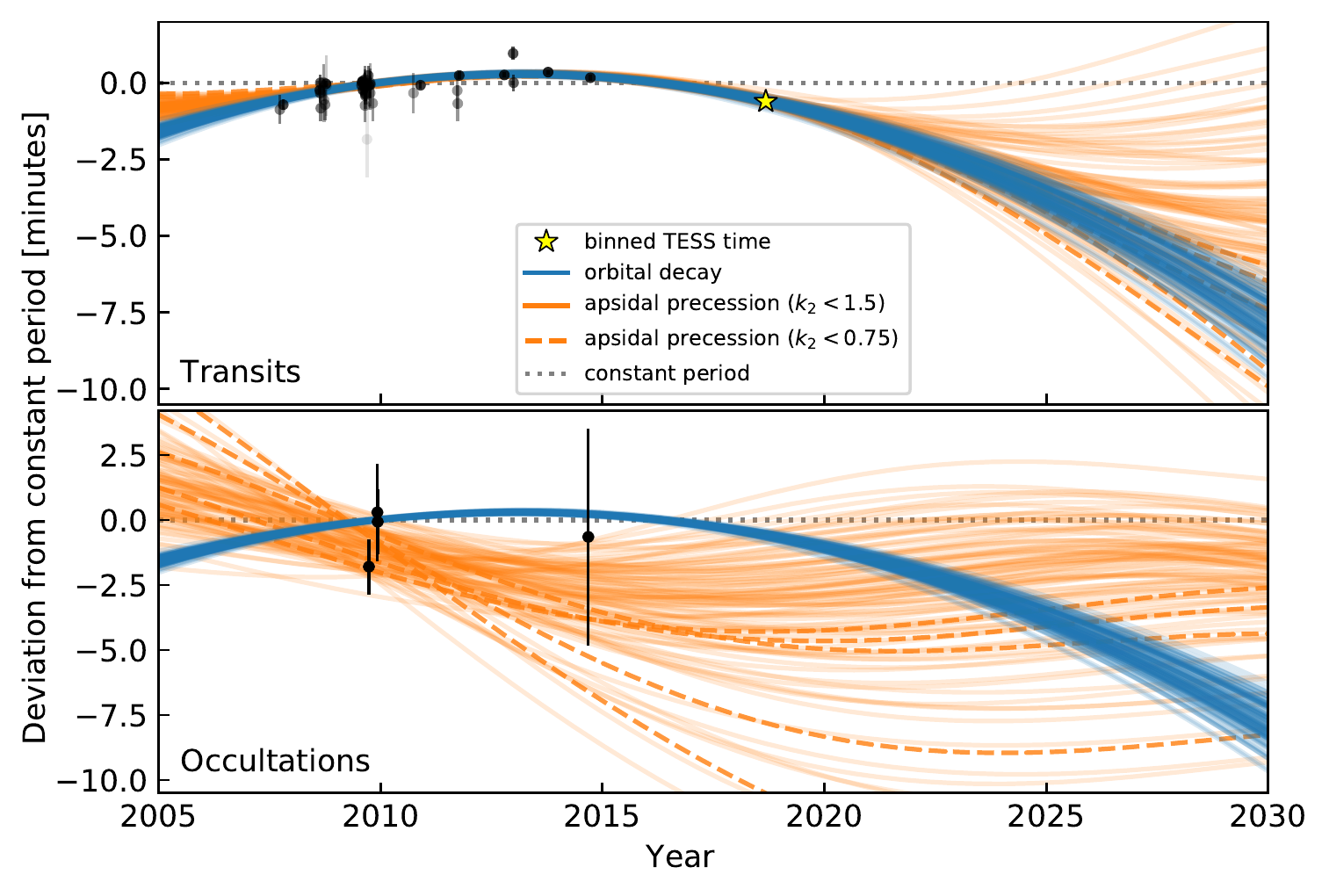}
	\end{center}
	\vspace{-0.7cm}
	\caption{
    {\bf Further observations are needed to confirm and understand
    the timing variations of WASP-4b.} \replaced{Dots}{Symbols} are as
    in Figure~\ref{fig:times}.  Lines are 100 random draws from the
    posteriors of the apsidal precession model (orange), and the
    orbital decay model (blue).  \added{The prior for the precession
    model assumes $k_{2,{\rm p}} \sim \mathcal{U}[0.015, 1.5]$; dashed orange
    lines emphasize samples with planetary Love numbers below 0.75}.    
		\label{fig:future}
	}
\end{figure*}

There are two \replaced{final}{other} small effects worth noting.  The
first is the \citet{shklovskii_possible_1970} effect due to the star's
proper motion, which leads to an apparent period change of $P\mu^2 d/
c$, which is only $6\times10^{-13}$ for the case of WASP-4.  The
second effect, described by \citet{rafikov_stellar_2009}, comes from
the star's on-sky motion altering our viewing angle, and leads to an
observed apsidal precession. The corresponding period change is
$\dot{P} \sim (P\mu)^2/2\pi$, which is on the order of $10^{-21}$ for
WASP-4, too small to be of any consequence.

%%%%%%%%%%%%%%%%%%%%%%%%%%%%%%%%%%%%%%%%%%
\section{Call for additional observations}
\label{sec:future}

A primary purpose of this work has been to call attention to the
timing anomaly of WASP-4 that has been sighted by TESS, and alert
observers to the need for follow-up transit timing, occultation
timing, and radial-velocity monitoring.  There is no unique
interpretation of the current data, and two of the possibilities ---
orbital decay and apsidal precession --- would be of great interest to
confirm. Detection of orbital decay would lead to an unusually direct
determination of a stellar dissipation rate. Detection of apsidal
precession would give a rare constraint on the interior density
distribution of an exoplanet.\added{ The third possibility --- a
massive outer companion --- would be the least exotic option, but
nonetheless a valuable discovery.}

If TESS is extended beyond its primary mission, it will likely observe
additional transits of WASP-4b in the early 2020s.
High-precision transit observations with
larger telescopes would also be useful.  In order to decide between
orbital decay and apsidal precession, occultation measurements in both
the near term and also in the mid-2020s will be needed 
(Figure~\ref{fig:future}).  More
radial-velocity data would help in the search for additional bodies
that could be causing dynamical perturbations, or an overall
acceleration of the host star.  The transit duration variations are
expected to be of order 10 seconds \citep{pal_periastron_2008}, and so
may remain out of reach.

TESS will also be monitoring the other known hot Jupiters, which will
reveal whether the timing anomalies seen in WASP-12 and now WASP-4 are
commonplace, and may shed some light on the circumstances in which
they arise. Other wide-field photometric surveys, such as the Next
Generation Transit Survey \citep{wheatley_next_2018}, HATPI
(\href{https://hatpi.org}{hatpi.org}) and PLATO
\citep{rauer_plato_2014} will also extend the time baseline of transit
timing for a large number of systems.

\acknowledgements
L.G.B.\ and J.N.W.\ gladly acknowledge helpful discussions with
A.~Bailey, J.~Goodman, K.~Patra\added{, D. Ragozzine,} and V.~Van
Eylen, and are grateful to the people who have turned TESS from an
idea into reality.  \added{L.G.B. thanks A.~Bixel and E.~May for
clarifying details concerning the available IMACS lightcurves.}
WASP-4 was included on the ``short-cadence'' target list thanks to the
Guest Investigator programs of J.\ Southworth and S.\ Kane (G011112
and G011183 respectively). 
J.N.W.\ thanks the TESS project and the Heising-Simons foundation for
supporting this work.
T.D. acknowledges support from MIT's Kavli Institute as a Kavli
postdoctoral fellow.
J.M.D. acknowledges funding from the European Research Council (ERC)
under the European Union's Horizon 2020 research and innovation
programme (grant agreement no. 679633; Exo-Atmos), and support from
the NWO TOP Grant Module 2 (Project Number 614.001.601).
J.E.R. was supported by the Harvard Future Faculty Leaders
Postdoctoral fellowship.
This paper includes data collected by the TESS mission, which are
publicly available from the Mikulski Archive for Space Telescopes
(MAST).
Funding for the TESS mission is provided by NASA's Science Mission
directorate.
This research has made use of the NASA Exoplanet Archive, which is
operated by the California Institute of Technology, under contract
with the National Aeronautics and Space Administration under the
Exoplanet Exploration Program.
This work made use of NASA's Astrophysics Data System Bibliographic
Services.
This research has made use of the VizieR catalogue access tool, CDS,
Strasbourg, France. The original description of the VizieR service was
published in A\&AS 143, 23.
This work has made use of data from the European Space Agency (ESA)
mission {\it Gaia} (\url{https://www.cosmos.esa.int/gaia}), processed
by the {\it Gaia} Data Processing and Analysis Consortium (DPAC,
\url{https://www.cosmos.esa.int/web/gaia/dpac/consortium}). Funding
for the DPAC has been provided by national institutions, in particular
the institutions participating in the {\it Gaia} Multilateral
Agreement.
\newline
\facility{
	TESS \citep{ricker_transiting_2015},
	Gaia \citep{gaia_collaboration_gaia_2016,gaia_collaboration_gaia_2018},
    Tycho-2 \citep{hog_tycho-2_2000},
    2MASS \citep{skrutskie_two_2006},
    APASS \citep{henden_data_2012},
    WISE \citep{wright_wide-field_2010}.
}
\software{
  \texttt{astrobase} \citep{bhatti_astrobase_2018},
  \texttt{astropy} \citep{the_astropy_collaboration_astropy_2018},
  \texttt{astroquery} \citep{astroquery_2018},
  \texttt{BATMAN} \citep{kreidberg_batman_2015},
  \texttt{corner} \citep{corner_2016},
  \texttt{emcee} \citep{foreman-mackey_emcee_2013},
  \texttt{IPython} \citep{perez_2007},
  \texttt{matplotlib} \citep{hunter_matplotlib_2007}, 
  \texttt{MESA} 
  \citep{paxton_modules_2011,paxton_modules_2013,paxton_modules_2015}
  \texttt{numpy} \citep{walt_numpy_2011}, 
  \texttt{pandas} \citep{mckinney-proc-scipy-2010},
  \texttt{radvel} \citep{fulton_radvel_2018},
  \texttt{scikit-learn} \citep{scikit-learn},
  \texttt{scipy} \citep{jones_scipy_2001}.
}

%\clearpage
%\newpage
\input{WASP-4b_transit_time_table.tex}

%\clearpage
%\newpage
\input{occultation_time_table.tex}

%\clearpage
%\newpage
\input{model_fit_table.tex}

\clearpage

\bibliographystyle{yahapj}                            
\bibliography{bibliography} 

%\clearpage
%\newpage

\appendix

%%%%%%%%%%%%%%%%%%%%%%%%%%%%%%%%%%%%%%%%%%
\section{Verifying the TESS Timestamps Using Other Hot Jupiters}
\label{sec:verify_tess}

\begin{figure*}[t!]
	\begin{center}
		\leavevmode
		\includegraphics[width=0.98\textwidth]{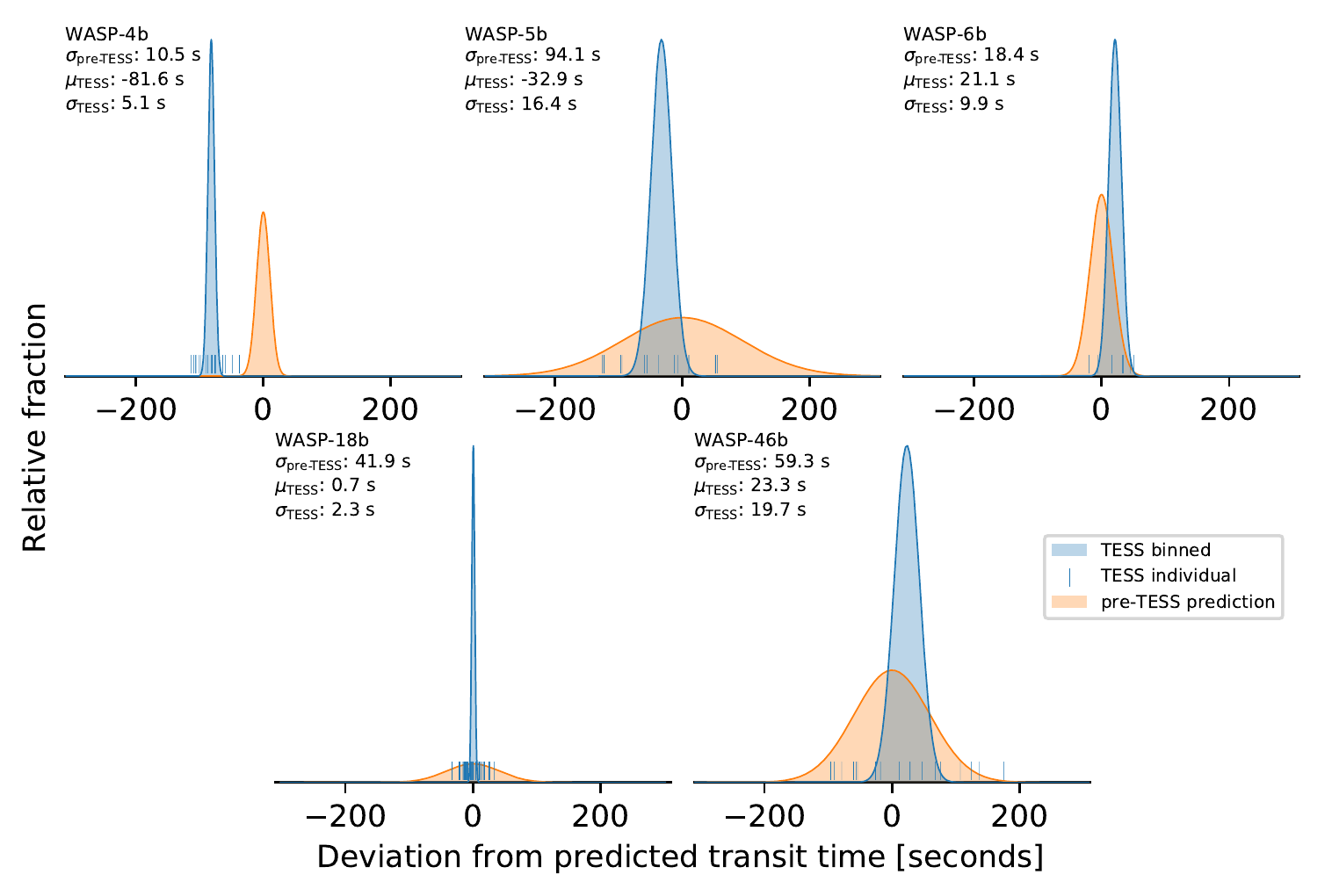}
	\end{center}
	\vspace{-0.5cm}
	\caption{
		{\bf There is no evidence for a systematic offset between TESS
			times and the barycentric reference.} While the WASP-4b transits
		fell about 82 seconds earlier than expected, other well-observed
		hot Jupiters, in particular WASP-6b and WASP-18b, arrived on time.
		Ticks are observed TESS transit midtimes; the orange distribution
		is a gaussian centered on zero with standard deviation
		(\replaced{$\sigma_{\rm predicted}$}{$\sigma_{\rm pre-TESS}$}) calculated from the pre-TESS transit
		times.  The blue distribution is a gaussian centered on the
		weighted average of the TESS times, with width equal to the
		uncertainty in the mean, {\it i.e.}, the standard deviation
		of the TESS residual times divided by $\sqrt{N-1}$, with $N$ the
		number of transits.
		\label{fig:hjs}
	}
\end{figure*}

An obvious concern that one might have about the WASP-4\added{b}
timing anomaly is that there might be a systematic offset between the
TESS time system and the time system in which the previous
observations have been reported.  There is a precedent for this type
of error: data from the Kepler mission was afflicted by a systematic
timing error that was corrected only late in the
mission~\citep[][Section 3.4]{kepler_DR19_2013}.

If the observed timing delay in WASP-4b were caused by a systematic
global offset between the TESS time system and the ${\rm BJD}_{\rm
TDB}$ reference, we would expect that it would be apparent in other
hot Jupiter systems, too. It would also be apparent in eclipsing
binary observations and any other periodic phenomena that have been
observed over a long time baseline. Here we examine only hot Jupiters
because of our greater familiarity with the data.

We repeated all the data reduction and analysis steps described in
this paper for other hot Jupiters observed by TESS for which timing
data exists spanning many years.  First, we checked which hot Jupiters
were observed over the first three TESS sectors using a combination of
\texttt{tessmaps}\footnote{\url{github.com/lgbouma/tessmaps}, 
	commit \texttt{569bbc2}} and
TEPCat \citep{southworth_homogeneous_2011}.  We recalculated the
barycentric corrections using the \citet{eastman_achieving_2010} code,
and found values that agreed with the lightcurve headers to within
about 1 second.  We then selected hot Jupiters for which there were at
least five distinct epochs reported in the peer-reviewed literature.
We required that each observation be of a single transit, that the
midpoint be fit as a free parameter, and that the time system be
clearly documented.  Our final hot Jupiter sample included WASP-4b,
5b, 6b, 18b, and 46b.  The collected and measured times are given in
Tables~5, 6, 7, and 8 for each.

We determined the best-fitting constant-period ephemeris based on the
pre-TESS data. Then we used the parameters and uncertainties in the
best-fitting model to calculate the predicted transit times during
the TESS observation period, as well as the uncertainty in the
predicted times.  The uncertainties are 11, 94, 18, 42, and 59 seconds
for WASP-4b, 5b, 6b, 18b, and 46b, respectively.  By comparing the
observed and predicted times, Figure~\ref{fig:hjs} shows that WASP-4b
is the only hot Jupiter that transited significantly earlier than
expected.

To use these results to place a quantitative limit on any global clock
offset, for each hot Jupiter we considered the model
\begin{equation}
  t_{\rm tra}(E) = t_0 + PE + t_{\rm offset},
\end{equation}
for $t_{\rm offset}$ a systematic constant offset between the reported
timestamps and the true ${\rm BJD}_{\rm TDB}$ reference.  Our priors
were
\begin{align}
  t_0 &\sim \mathcal{N}[t_0', \sigma_{t_0'}], \\
  P &\sim \mathcal{N}[P', \sigma_{P'}], \\
  t_{\rm offset} &\sim \mathcal{U}[-20\sigma_{t_0'},20\sigma_{t_0'}],
\end{align}
where $\mathcal{N}$ and $\mathcal{U}$ denote a normal and uniform
distribution, $(t_0', P')$ are the best-fit reference time and period
using only the pre-TESS transit times, and $(\sigma_{t_0'},
\sigma_{P'})$ are the corresponding uncertainties.

For each planet, we asked: what fraction of the posterior for $t_{\rm
offset}$ is consistent with an offset worse than $81.6$ seconds?  For
WASP-4b, the answer is unsurprisingly 50\%.  For WASP-6b, the most
constraining object, about 1 sample in 2 million is consistent with
such a timing offset ($4.9\sigma$).  For WASP-18b, 1 in 103 samples
would be consistent with this timing offset ($2.3\sigma$), and in
WASP-46b, the limit is 1 in 49 samples ($2.0\sigma$).  For WASP-5b,
the predicted time is too imprecise to rule out timing offsets at the
necessary amplitude.  Multiplying the three independent probabilities
for WASP-6b, 18b, and 46b, we can rule out $t_{\rm offset} < -81.6\
{\rm seconds}$ at $6.4\sigma$, or about about 1 part in 11 billion.

%\clearpage
%\newpage
\input{WASP-5b_transit_time_table.tex}
\input{WASP-6b_transit_time_table.tex}
\input{WASP-18b_transit_time_table.tex}
\input{WASP-46b_transit_time_table.tex}

%\clearpage
%\newpage

%\newpage

\listofchanges
\end{document}

%% file: WASP-4b_transit_time_table.tex
%% \begin{deluxetable}{} command tell LaTeX how many columns
%% there are and how to align them.
\startlongtable
\begin{deluxetable}{ccccc}
    
%% Keep a portrait orientation

%% Over-ride the default font size
%% Use Default (12pt)
\tabletypesize{\scriptsize}

%% Use \tablewidth{?pt} to over-ride the default table width.
%% If you are unhappy with the default look at the end of the
%% *.log file to see what the default was set at before adjusting
%% this value.

%% This is the title of the table.
\tablecaption{WASP-4b transit times, uncertainties, and references.}
\label{tab:transit_times}

%% This command over-rides LaTeX's natural table count
%% and replaces it with this number.  LaTeX will increment 
%% all other tables after this table based on this number
\tablenum{2}

%% The \tablehead gives provides the column headers.  It
%% is currently set up so that the column labels are on the
%% top line and the units surrounded by ()s are in the 
%% bottom line.  You may add more header information by writing
%% another line between these lines. For each column that requries
%% extra information be sure to include a \colhead{text} command
%% and remember to end any extra lines with \\ and include the 
%% correct number of &s.
\tablehead{
  \colhead{$t_{\rm tra}$ [BJD$_\mathrm{TDB}$]} &
  \colhead{$\sigma_{t_{\rm tra}}$ [days]} &
  \colhead{Epoch} & 
  \colhead{H13?} & 
  \colhead{Reference}
}

%% All data must appear between the \startdata and \enddata commands
% XXX pasted in from selected_transit_times.tex
\startdata
 2454368.59279 &      0.00033 &   -1073 &       1 &           \citet{wilson_wasp-4b_2008} \\
 2454396.69576 &      0.00012 &   -1052 &       1 &          \citet{gillon_improved_2009} \\
 2454697.79817 &      0.00009 &    -827 &       1 &             \citet{winn_transit_2009} \\
 2454701.81303 &      0.00018 &    -824 &       1 &             \citet{hoyer_tramos_2013} \\
 2454701.81280 &      0.00022 &    -824 &       1 &             \citet{hoyer_tramos_2013} \\
 2454705.82715 &      0.00029 &    -821 &       1 &             \citet{hoyer_tramos_2013} \\
 2454728.57767 &      0.00042 &    -804 &       1 &             \citet{hoyer_tramos_2013} \\
 2454732.59197 &      0.00050 &    -801 &       1 &             \citet{hoyer_tramos_2013} \\
 2454740.62125 &      0.00035 &    -795 &       1 &             \citet{hoyer_tramos_2013} \\
 2454748.65111 &      0.00007 &    -789 &       1 &             \citet{winn_transit_2009} \\
 2454752.66576 &      0.00069 &    -786 &       1 &           \citet{dragomir_terms_2011} \\
 2455041.72377 &      0.00018 &    -570 &       1 &             \citet{hoyer_tramos_2013} \\
 2455045.73853 &      0.00008 &    -567 &       1 &  \citet{sanchis-ojeda_starspots_2011} \\
 2455049.75325 &      0.00007 &    -564 &       1 &  \citet{sanchis-ojeda_starspots_2011} \\
 2455053.76774 &      0.00009 &    -561 &       1 &  \citet{sanchis-ojeda_starspots_2011} \\
 2455069.82661 &      0.00029 &    -549 &       1 &          \citet{nikolov_wasp-4b_2012} \\
 2455069.82617 &      0.00038 &    -549 &       1 &          \citet{nikolov_wasp-4b_2012} \\
 2455069.82670 &      0.00028 &    -549 &       1 &          \citet{nikolov_wasp-4b_2012} \\
 2455069.82676 &      0.00031 &    -549 &       1 &          \citet{nikolov_wasp-4b_2012} \\
 2455073.84108 &      0.00029 &    -546 &       1 &          \citet{nikolov_wasp-4b_2012} \\
 2455073.84128 &      0.00026 &    -546 &       1 &          \citet{nikolov_wasp-4b_2012} \\
 2455073.84111 &      0.00023 &    -546 &       1 &          \citet{nikolov_wasp-4b_2012} \\
 2455073.84114 &      0.00018 &    -546 &       1 &          \citet{nikolov_wasp-4b_2012} \\
 2455085.88418 &      0.00086 &    -537 &       1 &           \citet{dragomir_terms_2011} \\
 2455096.59148 &      0.00022 &    -529 &       1 &             \citet{hoyer_tramos_2013} \\
 2455100.60595 &      0.00012 &    -526 &       1 &  \citet{sanchis-ojeda_starspots_2011} \\
 2455112.64986 &      0.00039 &    -517 &       1 &          \citet{nikolov_wasp-4b_2012} \\
 2455112.65009 &      0.00033 &    -517 &       1 &          \citet{nikolov_wasp-4b_2012} \\
 2455112.65005 &      0.00031 &    -517 &       1 &          \citet{nikolov_wasp-4b_2012} \\
 2455112.65005 &      0.00049 &    -517 &       1 &          \citet{nikolov_wasp-4b_2012} \\
 2455132.72310 &      0.00041 &    -502 &       1 &             \citet{hoyer_tramos_2013} \\
 2455468.61943 &      0.00046 &    -251 &       1 &             \citet{hoyer_tramos_2013} \\
 2455526.16356 &      0.00008 &    -208 &       0 &       \citet{ranjan_atmospheric_2014} \\
 2455828.60375 &      0.00041 &      18 &       1 &             \citet{hoyer_tramos_2013} \\
 2455832.61815 &      0.00041 &      21 &       1 &             \citet{hoyer_tramos_2013} \\
 2455844.66287 &      0.00009 &      30 &       0 &           \citet{huitson_gemini_2017} \\
 2456216.69123 &      0.00006 &     308 &       0 &           \citet{huitson_gemini_2017} \\
 2456288.95622 &      0.00015 &     362 &       0 &              Baxter et al.\ (in prep) \\
 2456292.97025 &      0.00019 &     365 &       0 &              Baxter et al.\ (in prep) \\
 2456576.67556 &      0.00005 &     577 &       0 &           \citet{huitson_gemini_2017} \\
 2456924.61561 &      0.00006 &     837 &       0 &           \citet{huitson_gemini_2017} \\
 2458355.18490 &      0.00025 &    1906 &       0 &                             This work \\
 2458356.52251 &      0.00027 &    1907 &       0 &                             This work \\
 2458357.86105 &      0.00026 &    1908 &       0 &                             This work \\
 2458359.19946 &      0.00026 &    1909 &       0 &                             This work \\
 2458360.53707 &      0.00028 &    1910 &       0 &                             This work \\
 2458361.87538 &      0.00025 &    1911 &       0 &                             This work \\
 2458363.21411 &      0.00027 &    1912 &       0 &                             This work \\
 2458364.55193 &      0.00025 &    1913 &       0 &                             This work \\
 2458365.89057 &      0.00026 &    1914 &       0 &                             This work \\
 2458369.90506 &      0.00028 &    1917 &       0 &                             This work \\
 2458371.24298 &      0.00026 &    1918 &       0 &                             This work \\
 2458372.58124 &      0.00026 &    1919 &       0 &                             This work \\
 2458373.91981 &      0.00028 &    1920 &       0 &                             This work \\
 2458375.25792 &      0.00025 &    1921 &       0 &                             This work \\
 2458376.59623 &      0.00024 &    1922 &       0 &                             This work \\
 2458377.93434 &      0.00026 &    1923 &       0 &                             This work \\
 2458379.27319 &      0.00025 &    1924 &       0 &                             This work \\
 2458380.61098 &      0.00028 &    1925 &       0 &                             This work \\
\enddata

%% Include any \tablenotetext{key}{text}, \tablerefs{ref list},
%% or \tablecomments{text} between the \enddata and 
%% \end{deluxetable} commands

%% General table comment marker
\tablecomments{
  $t_{\rm tra}$ is the measured transit midtime, and $\sigma_{t_{\rm
  tra}}$ is its $1\sigma$ uncertainty.
  \replaced{$\sigma_{t_0}$}{$\sigma_{t_{\rm tra}}$} was evaluated from the 
  sampled posteriors by taking
  the maximum of the difference between the 84th percentile
  minus the median, and the median minus the 16th percentile.
  \added{The resulting error variances then appeared to have been overestimated, so 
  we lowered the uncertainties as described in \S~\ref{sec:measurement}.}
  The ``Reference'' column refers to the work describing the
  original observations.
  The ``H13?'' column is 1 if the mid-time value was taken from
  \citet{hoyer_tramos_2013}.  Otherwise, the mid-time came from the
  column listed in ``Reference''.
  The \citealt{hoyer_tramos_2013} BJD$_{\rm TT}$ times are equal to
  BJD$_{\rm TDB}$ for our purposes \citep{urban_explanatory_2012}.
  We omitted the timing measurements from
  \citet{southworth_high-precision_2009}, since there were technical
  problems with the computer clock at the time of
  observation~\citep{nikolov_wasp-4b_2012}.
  The two Baxter et al.\ (in prep) times were obtained from
  Spitzer/IRAC transit light curves in the 3.6$\mu$m and 4.5$\mu$m
  channels.
}
\end{deluxetable}

%% file: occultation_time_table.tex
%% \begin{deluxetable}{} command tell LaTeX how many columns
%% there are and how to align them.
\startlongtable
\begin{deluxetable}{cccc}
    
%% Keep a portrait orientation

%% Over-ride the default font size
%% Use Default (12pt)
\tabletypesize{\footnotesize}

%% Use \tablewidth{?pt} to over-ride the default table width.
%% If you are unhappy with the default look at the end of the
%% *.log file to see what the default was set at before adjusting
%% this value.

%% This is the title of the table.
\tablecaption{WASP-4b occultation times, uncertainties, and references.}
\label{tab:occultation_times}

%% This command over-rides LaTeX's natural table count
%% and replaces it with this number.  LaTeX will increment 
%% all other tables after this table based on this number
\tablenum{3}

%% The \tablehead gives provides the column headers.  It
%% is currently set up so that the column labels are on the
%% top line and the units surrounded by ()s are in the 
%% bottom line.  You may add more header information by writing
%% another line between these lines. For each column that requries
%% extra information be sure to include a \colhead{text} command
%% and remember to end any extra lines with \\ and include the 
%% correct number of &s.
\tablehead{
  \colhead{$t_{\rm occ}$ [BJD$_\mathrm{TDB}$]} &
  \colhead{$\sigma_{t_{\rm occ}}$ [days]} &
  \colhead{Epoch} & 
  \colhead{Reference}
}

%% All data must appear between the \startdata and \enddata commands
% XXX pasted in from selected_transit_times.tex
\startdata
 2455102.61210 &      0.00074 &    -511 &  \citet{caceres_ground-based_2011}\tablenotemark{a} \\
 2455172.20159 &      0.00130 &    -459 &      \citet{beerer_secondary_2011} \\
 2455174.87780 &      0.00087 &    -457 &      \citet{beerer_secondary_2011} \\
 2456907.88714 &      0.00290 &     838 &        \citet{zhou_secondary_2015}\tablenotemark{b} \\
\enddata

%% Include any \tablenotetext{key}{text}, \tablerefs{ref list},
%% or \tablecomments{text} between the \enddata and 
%% \end{deluxetable} commands

%% General table comment marker
\tablecomments{
	$t_{\rm occ}$ is the measured occultation midtime, minus the
	$2a/c=22.8$ second light travel time;
	$\sigma_{t_{\rm occ}}$ is the $1\sigma$ uncertainty on the occultation
	time.
}
\tablenotetext{a}{
\citet{caceres_ground-based_2011} reported this time in ``HJD'', with
an unspecified time standard. We assumed the time was originally in
${\rm HJD}_{\rm UTC}$, and converted to ${\rm BJD}_{\rm TDB}$ for the
tabulated time.
}
\tablenotetext{b}{
\citet{zhou_secondary_2015} fixed the epoch, and let $e\cos\omega$
float. Using the reported dates of observation, we converted their
$e\cos\omega$ values into an occultation time using
Equation~\ref{eq:occultation_time} of the text. 
}

\end{deluxetable}

%% file: model_fit_table.tex
%\renewcommand{\arraystretch}{1.0}

\startlongtable
\begin{deluxetable}{lc}

\tabletypesize{\footnotesize}

\tablenum{4}

%\tablewidth{0pt}

\tablecaption{Best-fit transit timing model parameters.}
\label{tab:bestfit}

\tablehead{
  \colhead{Parameter} &
  \colhead{Median Value~(Unc.)\tablenotemark{a}}
}

\startdata
~~~~~~{\it Constant period} &  \\
$t_0$\,[${\rm BJD}_{\rm TBD}$]    & 2455804.515752(+19)(-19)              \\
$P$\,[days]                       & 1.338231466(+23)(-22)                 \\
~~~~~~{\it Constant period derivative} &  \\
$t_0$~[${\rm BJD}_{\rm TBD}$]     & 2455804.515918(+24)(-24)              \\
$P$\,[days]                       & 1.338231679(+31)(-31)                 \\
$dP/dt$                           & $-4.00(+37)(-38) \times 10^{-10}$     \\
~~~~~~{\it Apsidal precession (wide prior)} &  \\
$t_0$~[${\rm BJD}_{\rm TBD}$]     & 2455804.51530(+25)(-31)               \\
$P_{\rm s}$\,[days]               & 1.33823127(+20)(-48)                  \\
$e$                               & $1.92^{+1.93}_{-0.76} \times 10^{-3}$ \\
$\omega_0$\,[rad]                 & 2.40(+38)(-34)                        \\
$d\omega/dE$~[rad\,epoch$^{-1}$]  & $8.70^{+3.01}_{-2.30} \times 10^{-4}$ \\
\enddata
\tablenotetext{a}{
The numbers in parenthesis give the $68\%$ confidence interval for the final
two digits, where appropriate.
}
\end{deluxetable}

%% file: WASP-5b_transit_time_table.tex
%% \begin{deluxetable}{} command tell LaTeX how many columns
%% there are and how to align them.
\startlongtable
\begin{deluxetable}{ccccc}
    
%% Keep a portrait orientation

%% Over-ride the default font size
%% Use Default (12pt)
\tabletypesize{\scriptsize}

%% Use \tablewidth{?pt} to over-ride the default table width.
%% If you are unhappy with the default look at the end of the
%% *.log file to see what the default was set at before adjusting
%% this value.

%% This is the title of the table.
\tablecaption{WASP-5b transit times, uncertainties, and references.}
\label{tab:WASP-5b}

%% This command over-rides LaTeX's natural table count
%% and replaces it with this number.  LaTeX will increment 
%% all other tables after this table based on this number
\tablenum{5}

%% The \tablehead gives provides the column headers.  It
%% is currently set up so that the column labels are on the
%% top line and the units surrounded by ()s are in the 
%% bottom line.  You may add more header information by writing
%% another line between these lines. For each column that requries
%% extra information be sure to include a \colhead{text} command
%% and remember to end any extra lines with \\ and include the 
%% correct number of &s.
\tablehead{
  \colhead{$t_{\rm tra}$ [BJD$_\mathrm{TDB}$]} &
  \colhead{$\sigma_{t_{\rm tra}}$ [days]} &
  \colhead{Epoch} & 
  \colhead{Reference}
}

%% All data must appear between the \startdata and \enddata commands
\startdata
 2454383.76750 &      0.00040 &    -885 &           \citet{anderson_wasp-5b_2008} \\
 2454387.02275 &      0.00100 &    -883 &           \citet{anderson_wasp-5b_2008} \\
 2454636.17459 &      0.00082 &    -730 &         \citet{fukui_measurements_2011} \\
 2454699.68303 &      0.00041 &    -691 &              \citet{hoyer_transit_2012} \\
 2454707.82465 &      0.00052 &    -686 &              \citet{hoyer_transit_2012} \\
 2454707.82523 &      0.00025 &    -686 &  \citet{southworth_high-precision_2009} \\
 2454730.62243 &      0.00031 &    -672 &  \citet{southworth_high-precision_2009} \\
 2454730.62301 &      0.00076 &    -672 &              \citet{hoyer_transit_2012} \\
 2454761.56356 &      0.00047 &    -653 &              \citet{hoyer_transit_2012} \\
 2454772.96212 &      0.00075 &    -646 &         \citet{fukui_measurements_2011} \\
 2454774.59093 &      0.00030 &    -645 &              \citet{hoyer_transit_2012} \\
 2454787.61792 &      0.00069 &    -637 &              \citet{hoyer_transit_2012} \\
 2455005.82714 &      0.00036 &    -503 &              \citet{hoyer_transit_2012} \\
 2455049.79540 &      0.00080 &    -476 &              \citet{hoyer_transit_2012} \\
 2455075.84947 &      0.00056 &    -460 &             \citet{dragomir_terms_2011} \\
 2455079.10830 &      0.00079 &    -458 &         \citet{fukui_measurements_2011} \\
 2455110.04607 &      0.00089 &    -439 &         \citet{fukui_measurements_2011} \\
 2455123.07611 &      0.00079 &    -431 &         \citet{fukui_measurements_2011} \\
 2455129.58759 &      0.00043 &    -427 &              \citet{hoyer_transit_2012} \\
 2455364.08150 &      0.00110 &    -283 &         \citet{fukui_measurements_2011} \\
 2455377.10955 &      0.00093 &    -275 &         \citet{fukui_measurements_2011} \\
 2455448.75927 &      0.00110 &    -231 &             \citet{dragomir_terms_2011} \\
 2456150.61479 &      0.00056 &     200 &          \citet{moyano_multi-band_2017} \\
 2456150.61396 &      0.00057 &     200 &          \citet{moyano_multi-band_2017} \\
 2458355.50829 &      0.00083 &    1554 &                               This work \\
 2458357.13741 &      0.00071 &    1555 &                               This work \\
 2458358.76412 &      0.00068 &    1556 &                               This work \\
 2458360.39377 &      0.00070 &    1557 &                               This work \\
 2458362.02273 &      0.00073 &    1558 &                               This work \\
 2458363.64908 &      0.00090 &    1559 &                               This work \\
 2458365.27827 &      0.00071 &    1560 &                               This work \\
 2458366.90627 &      0.00075 &    1561 &                               This work \\
 2458370.16411 &      0.00076 &    1563 &                               This work \\
 2458371.79126 &      0.00071 &    1564 &                               This work \\
 2458373.42123 &      0.00075 &    1565 &                               This work \\
 2458375.04910 &      0.00069 &    1566 &                               This work \\
 2458376.67856 &      0.00074 &    1567 &                               This work \\
 2458378.30530 &      0.00087 &    1568 &                               This work \\
 2458379.93419 &      0.00082 &    1569 &                               This work \\
\enddata

%% Include any \tablenotetext{key}{text}, \tablerefs{ref list},
%% or \tablecomments{text} between the \enddata and 
%% \end{deluxetable} commands

%% General table comment marker
\tablecomments{
    $t_{\rm tra}$ is the measured transit midtime, and $\sigma_{t_{\rm tra}}$ is its
    $1\sigma$ uncertainty.
    The ``Reference'' column refers to the work describing the
    original observations.
    All the literature times except for the two \citet{moyano_multi-band_2017}
    times are from the homogeneous \citet{hoyer_transit_2012} analysis.
}

\end{deluxetable}

%% file: WASP-6b_transit_time_table.tex
%% \begin{deluxetable}{} command tell LaTeX how many columns
%% there are and how to align them.
\startlongtable
\begin{deluxetable}{ccccc}
    
%% Keep a portrait orientation

%% Over-ride the default font size
%% Use Default (12pt)
\tabletypesize{\scriptsize}
%% Use \tablewidth{?pt} to over-ride the default table width.
%% If you are unhappy with the default look at the end of the
%% *.log file to see what the default was set at before adjusting
%% this value.

%% This is the title of the table.
\tablecaption{WASP-6b transit times, uncertainties, and references.}
\label{tab:WASP-6b}

%% This command over-rides LaTeX's natural table count
%% and replaces it with this number.  LaTeX will increment 
%% all other tables after this table based on this number
\tablenum{6}

%% The \tablehead gives provides the column headers.  It
%% is currently set up so that the column labels are on the
%% top line and the units surrounded by ()s are in the 
%% bottom line.  You may add more header information by writing
%% another line between these lines. For each column that requries
%% extra information be sure to include a \colhead{text} command
%% and remember to end any extra lines with \\ and include the 
%% correct number of &s.
\tablehead{
  \colhead{$t_{\rm tra}$ [BJD$_\mathrm{TDB}$]} &
  \colhead{$\sigma_{t_{\rm tra}}$ [days]} &
  \colhead{Epoch} & 
  \colhead{Reference}
}

%% All data must appear between the \startdata and \enddata commands
\startdata
 2454425.02167 &      0.00022 &    -398 &        \citet{gillon_discovery_2009} \\
 2455009.83622 &      0.00021 &    -224 &  \citet{tregloan-reed_transits_2015} \\
 2455046.80720 &      0.00015 &    -213 &  \citet{tregloan-reed_transits_2015} \\
 2455073.69529 &      0.00013 &    -205 &  \citet{tregloan-reed_transits_2015} \\
 2455409.79541 &      0.00010 &    -105 &  \citet{tregloan-reed_transits_2015} \\
 2455446.76621 &      0.00058 &     -94 &          \citet{dragomir_terms_2011} \\
 2455473.65439 &      0.00097 &     -86 &     \citet{jordan_ground-based_2013} \\
 2455846.72540 &      0.00045 &      25 &         \citet{sada_extrasolar_2012} \\
 2456088.71801 &      0.00013 &      97 &             \citet{nikolov_hst_2015} \\
 2456095.43974 &      0.00017 &      99 &             \citet{nikolov_hst_2015} \\
 2456132.41082 &      0.00017 &     110 &             \citet{nikolov_hst_2015} \\
 2458357.39410 &      0.00033 &     772 &                            This work \\
 2458360.75573 &      0.00033 &     773 &                            This work \\
 2458364.11691 &      0.00032 &     774 &                            This work \\
 2458370.83872 &      0.00033 &     776 &                            This work \\
 2458374.19952 &      0.00031 &     777 &                            This work \\
 2458377.56026 &      0.00033 &     778 &                            This work \\
 2458380.92185 &      0.00038 &     779 &                            This work \\
\enddata

%% Include any \tablenotetext{key}{text}, \tablerefs{ref list},
%% or \tablecomments{text} between the \enddata and 
%% \end{deluxetable} commands

%% General table comment marker
\tablecomments{
    $t_{\rm tra}$ is the measured transit midtime, and $\sigma_{t_{\rm tra}}$ is its
    $1\sigma$ uncertainty.
    The ``Reference'' column refers to the work describing the
    original observations.
}

\end{deluxetable}

%% file: WASP-18b_transit_time_table.tex
%% \begin{deluxetable}{} command tell LaTeX how many columns
%% there are and how to align them.
\startlongtable
\begin{deluxetable}{ccccc}
    
%% Keep a portrait orientation

%% Over-ride the default font size
%% Use Default (12pt)
\tabletypesize{\scriptsize}
%% Use \tablewidth{?pt} to over-ride the default table width.
%% If you are unhappy with the default look at the end of the
%% *.log file to see what the default was set at before adjusting
%% this value.

%% This is the title of the table.
\tablecaption{WASP-18b transit times, uncertainties, and references.}
\label{tab:WASP-18b}

%% This command over-rides LaTeX's natural table count
%% and replaces it with this number.  LaTeX will increment 
%% all other tables after this table based on this number
\tablenum{7}

%% The \tablehead gives provides the column headers.  It
%% is currently set up so that the column labels are on the
%% top line and the units surrounded by ()s are in the 
%% bottom line.  You may add more header information by writing
%% another line between these lines. For each column that requries
%% extra information be sure to include a \colhead{text} command
%% and remember to end any extra lines with \\ and include the 
%% correct number of &s.
\tablehead{
  \colhead{$t_{\rm tra}$ [BJD$_\mathrm{TDB}$]} &
  \colhead{$\sigma_{t_{\rm tra}}$ [days]} &
  \colhead{Epoch} & 
  \colhead{Reference}
}

%% All data must appear between the \startdata and \enddata commands
\startdata
 2454221.48163 &      0.00038 &   -4037 &    \citet{hellier_orbital_2009} \\
 2455221.30420 &      0.00010 &   -2975 &     \citet{maxted_spitzer_2013} \\
 2455432.18970 &      0.00010 &   -2751 &     \citet{maxted_spitzer_2013} \\
 2455470.78850 &      0.00040 &   -2710 &     \citet{maxted_spitzer_2013} \\
 2455473.61440 &      0.00090 &   -2707 &     \citet{maxted_spitzer_2013} \\
 2455554.57860 &      0.00050 &   -2621 &     \citet{maxted_spitzer_2013} \\
 2455570.58400 &      0.00048 &   -2604 &     \citet{maxted_spitzer_2013} \\
 2455876.55590 &      0.00130 &   -2279 &     \citet{maxted_spitzer_2013} \\
 2456896.14780 &      0.00080 &   -1196 &  \citet{wilkins_searching_2017} \\
 2457255.78320 &      0.00030 &    -814 &  \citet{wilkins_searching_2017} \\
 2457319.80100 &      0.00039 &    -746 &  \citet{wilkins_searching_2017} \\
 2458354.45782 &      0.00016 &     353 &                       This work \\
 2458355.39933 &      0.00015 &     354 &                       This work \\
 2458356.34070 &      0.00018 &     355 &                       This work \\
 2458357.28229 &      0.00018 &     356 &                       This work \\
 2458358.22348 &      0.00018 &     357 &                       This work \\
 2458359.16523 &      0.00020 &     358 &                       This work \\
 2458360.10661 &      0.00017 &     359 &                       This work \\
 2458361.04810 &      0.00017 &     360 &                       This work \\
 2458361.98968 &      0.00016 &     361 &                       This work \\
 2458362.93130 &      0.00018 &     362 &                       This work \\
 2458363.87267 &      0.00018 &     363 &                       This work \\
 2458364.81374 &      0.00017 &     364 &                       This work \\
 2458365.75525 &      0.00019 &     365 &                       This work \\
 2458366.69709 &      0.00018 &     366 &                       This work \\
 2458369.52128 &      0.00017 &     369 &                       This work \\
 2458370.46281 &      0.00017 &     370 &                       This work \\
 2458371.40407 &      0.00017 &     371 &                       This work \\
 2458372.34537 &      0.00018 &     372 &                       This work \\
 2458373.28728 &      0.00018 &     373 &                       This work \\
 2458374.22818 &      0.00016 &     374 &                       This work \\
 2458375.16977 &      0.00017 &     375 &                       This work \\
 2458376.11132 &      0.00018 &     376 &                       This work \\
 2458377.05267 &      0.00017 &     377 &                       This work \\
 2458377.99444 &      0.00018 &     378 &                       This work \\
 2458378.93573 &      0.00016 &     379 &                       This work \\
 2458379.87722 &      0.00017 &     380 &                       This work \\
 2458380.81889 &      0.00018 &     381 &                       This work \\
 2458386.46729 &      0.00016 &     387 &                       This work \\
 2458387.40888 &      0.00017 &     388 &                       This work \\
 2458388.35021 &      0.00016 &     389 &                       This work \\
 2458389.29161 &      0.00015 &     390 &                       This work \\
 2458390.23334 &      0.00016 &     391 &                       This work \\
 2458391.17452 &      0.00016 &     392 &                       This work \\
 2458392.11593 &      0.00016 &     393 &                       This work \\
 2458393.05748 &      0.00015 &     394 &                       This work \\
 2458393.99898 &      0.00016 &     395 &                       This work \\
 2458394.94024 &      0.00017 &     396 &                       This work \\
 2458396.82309 &      0.00015 &     398 &                       This work \\
 2458397.76450 &      0.00015 &     399 &                       This work \\
 2458398.70656 &      0.00016 &     400 &                       This work \\
 2458399.64748 &      0.00015 &     401 &                       This work \\
 2458399.64748 &      0.00015 &     401 &                       This work \\
 2458400.58898 &      0.00017 &     402 &                       This work \\
 2458401.53083 &      0.00016 &     403 &                       This work \\
 2458402.47209 &      0.00017 &     404 &                       This work \\
 2458403.41360 &      0.00016 &     405 &                       This work \\
 2458404.35492 &      0.00017 &     406 &                       This work \\
\enddata

%% Include any \tablenotetext{key}{text}, \tablerefs{ref list},
%% or \tablecomments{text} between the \enddata and 
%% \end{deluxetable} commands

%% General table comment marker
\tablecomments{
    $t_{\rm tra}$ is the measured transit midtime, and $\sigma_{t_{\rm tra}}$ is its
    $1\sigma$ uncertainty.
    The ``Reference'' column refers to the work describing the
    original observations.
    All the literature times are from the homogeneous
    \citet{wilkins_searching_2017} analysis.
}

\end{deluxetable}

%% file: WASP-46b_transit_time_table.tex
%% \begin{deluxetable}{} command tell LaTeX how many columns
%% there are and how to align them.
\startlongtable
\begin{deluxetable}{ccccc}
    
%% Keep a portrait orientation

%% Over-ride the default font size
%% Use Default (12pt)
\tabletypesize{\scriptsize}
%% Use \tablewidth{?pt} to over-ride the default table width.
%% If you are unhappy with the default look at the end of the
%% *.log file to see what the default was set at before adjusting
%% this value.

%% This is the title of the table.
\tablecaption{WASP-46b transit times, uncertainties, and references.}
\label{tab:WASP-46b}

%% This command over-rides LaTeX's natural table count
%% and replaces it with this number.  LaTeX will increment 
%% all other tables after this table based on this number
\tablenum{8}

%% The \tablehead gives provides the column headers.  It
%% is currently set up so that the column labels are on the
%% top line and the units surrounded by ()s are in the 
%% bottom line.  You may add more header information by writing
%% another line between these lines. For each column that requries
%% extra information be sure to include a \colhead{text} command
%% and remember to end any extra lines with \\ and include the 
%% correct number of &s.
\tablehead{
  \colhead{$t_{\rm tra}$ [BJD$_\mathrm{TDB}$]} &
  \colhead{$\sigma_{t_{\rm tra}}$ [days]} &
  \colhead{Epoch} & 
  \colhead{Reference}
}

%% All data must appear between the \startdata and \enddata commands
\startdata
 2455396.60785 &      0.00062 &    -673 &  \citet{anderson_wasp-44b_2012} \\
 2455449.53082 &      0.00026 &    -636 &  \citet{anderson_wasp-44b_2012} \\
 2455722.73178 &      0.00023 &    -445 &    \citet{ciceri_physical_2016} \\
 2455757.06195 &      0.00094 &    -421 &    \citet{petrucci_search_2018} \\
 2455858.61833 &      0.00009 &    -350 &    \citet{ciceri_physical_2016} \\
 2456108.92771 &      0.00094 &    -175 &    \citet{petrucci_search_2018} \\
 2456111.79422 &      0.00016 &    -173 &    \citet{ciceri_physical_2016} \\
 2456111.79413 &      0.00012 &    -173 &    \citet{ciceri_physical_2016} \\
 2456111.79424 &      0.00015 &    -173 &    \citet{ciceri_physical_2016} \\
 2456130.38895 &      0.00042 &    -160 &    \citet{petrucci_search_2018} \\
 2456131.81456 &      0.00112 &    -159 &    \citet{petrucci_search_2018} \\
 2456194.75916 &      0.00027 &    -115 &    \citet{ciceri_physical_2016} \\
 2456217.64127 &      0.00015 &     -99 &    \citet{ciceri_physical_2016} \\
 2456217.64156 &      0.00013 &     -99 &    \citet{ciceri_physical_2016} \\
 2456227.65574 &      0.00060 &     -92 &    \citet{petrucci_search_2018} \\
 2456407.88096 &      0.00015 &      34 &    \citet{ciceri_physical_2016} \\
 2456407.88085 &      0.00018 &      34 &    \citet{ciceri_physical_2016} \\
 2456407.88148 &      0.00028 &      34 &    \citet{ciceri_physical_2016} \\
 2456407.88159 &      0.00043 &      34 &    \citet{ciceri_physical_2016} \\
 2456460.80526 &      0.00017 &      71 &    \citet{ciceri_physical_2016} \\
 2456460.80450 &      0.00024 &      71 &    \citet{ciceri_physical_2016} \\
 2456460.80547 &      0.00064 &      71 &    \citet{ciceri_physical_2016} \\
 2456510.86818 &      0.00060 &     106 &    \citet{petrucci_search_2018} \\
 2456510.86699 &      0.00015 &     106 &    \citet{petrucci_search_2018} \\
 2456516.58667 &      0.00119 &     110 &    \citet{petrucci_search_2018} \\
 2456520.88012 &      0.00064 &     113 &    \citet{petrucci_search_2018} \\
 2456533.75260 &      0.00071 &     122 &    \citet{ciceri_physical_2016} \\
 2456533.75480 &      0.00015 &     122 &    \citet{ciceri_physical_2016} \\
 2456576.66289 &      0.00109 &     152 &    \citet{petrucci_search_2018} \\
 2456589.54197 &      0.00090 &     161 &    \citet{petrucci_search_2018} \\
 2456609.56653 &      0.00043 &     175 &    \citet{petrucci_search_2018} \\
 2456839.85440 &      0.00123 &     336 &    \citet{petrucci_search_2018} \\
 2456862.74085 &      0.00048 &     352 &    \citet{petrucci_search_2018} \\
 2456882.76566 &      0.00073 &     366 &    \citet{petrucci_search_2018} \\
 2456885.62429 &      0.00053 &     368 &    \citet{petrucci_search_2018} \\
 2456915.66040 &      0.00123 &     389 &    \citet{petrucci_search_2018} \\
 2456942.83880 &      0.00078 &     408 &    \citet{petrucci_search_2018} \\
 2456948.56384 &      0.00074 &     412 &    \citet{petrucci_search_2018} \\
 2457274.68458 &      0.00184 &     640 &    \citet{petrucci_search_2018} \\
 2457294.70886 &      0.00140 &     654 &    \citet{petrucci_search_2018} \\
 2457550.74797 &      0.00031 &     833 &    \citet{petrucci_search_2018} \\
 2457593.65692 &      0.00024 &     863 &    \citet{petrucci_search_2018} \\
 2457600.80985 &      0.00039 &     868 &    \citet{petrucci_search_2018} \\
 2457610.82286 &      0.00020 &     875 &    \citet{petrucci_search_2018} \\
 2458326.00972 &      0.00091 &    1375 &                       This work \\
 2458327.43899 &      0.00093 &    1376 &                       This work \\
 2458328.86970 &      0.00094 &    1377 &                       This work \\
 2458330.29965 &      0.00105 &    1378 &                       This work \\
 2458331.73234 &      0.00105 &    1379 &                       This work \\
 2458333.15977 &      0.00086 &    1380 &                       This work \\
 2458334.59230 &      0.00095 &    1381 &                       This work \\
 2458336.02222 &      0.00082 &    1382 &                       This work \\
 2458337.45111 &      0.00099 &    1383 &                       This work \\
 2458340.31143 &      0.00093 &    1385 &                       This work \\
 2458341.74347 &      0.00093 &    1386 &                       This work \\
 2458343.17362 &      0.00093 &    1387 &                       This work \\
 2458344.60303 &      0.00110 &    1388 &                       This work \\
 2458346.03436 &      0.00091 &    1389 &                       This work \\
 2458347.46335 &      0.00168 &    1390 &                       This work \\
 2458348.89621 &      0.00086 &    1391 &                       This work \\
 2458350.32672 &      0.00101 &    1392 &                       This work \\
 2458351.75486 &      0.00103 &    1393 &                       This work \\
\enddata

%% Include any \tablenotetext{key}{text}, \tablerefs{ref list},
%% or \tablecomments{text} between the \enddata and 
%% \end{deluxetable} commands

%% General table comment marker
\tablecomments{
    $t_{\rm tra}$ is the measured transit midtime, and $\sigma_{t_{\rm tra}}$ is its
    $1\sigma$ uncertainty.
    The ``Reference'' column refers to the work describing the
    original observations.
    All the literature times are from the homogeneous
    \citet{petrucci_search_2018} analysis. 14 of the lightcurves
    were acquired by ETD observers \citep[see][]{petrucci_search_2018}.
}

\end{deluxetable}